\documentclass[10pt]{article}
\usepackage{amsfonts}
\usepackage{amsmath}

\begin{document}

\title{Rich gauge structures from an unitary approach of some massless gauge fields of spins one and two}

\author{Eugen-Mihaita CIOROIANU \\
Department of Physics, University of Craiova\\
13 Al. I. Cuza Str., Craiova 200585, Romania\\
manache@central.ucv.ro}
\maketitle

\begin{abstract}
Considering a bosonic ($1$-)form-valued $k$-form with a second-order
Lagrangian dynamics [depending on two arbitrary real constants] we firstly
perform the Dirac analysis. The procedure implies a partition of cardinally
seven for the plane of the real parameters that label the starting
Lagrangian. In each of the seven partition's components one determines the
number and the nature of independent degrees of freedom and also a
generating set of gauge transformations. Secondly, with the help of some
auxiliary gauge/matter tensor gauge fields, in each of the seven situations,
we construct the first-order Lagrangian density corresponding to the
second-order one.
\end{abstract}


\section{Introduction}

The main building blocks of fundamental interactions consist in particles of
spins one and two. These play the role of quanta for the massless gauge
fields of spins one and two that mediate all the fundamental interactions in
Nature. Due to this status of the spin-$1$ and spin-$2$ gauge fields, there
appear the natural questions: i) \emph{is it possible to treat them in an
unifying manner?} and ii) if the answer is positive \emph{what benefits
brings this unification?} This problem was the basic clue in an old attempt
to unify gravity with electromagnetism [in four spacetime dimensions]
proposed by Einstein and developed by himself \cite{ein} and others \cite%
{sch}, \cite{moff}. In the present paper we shall prove that the answer to
the first question is alway \emph{positive in any spacetime dimension
[greater or equal to four] at the level of free fields}. In order to do
this, we consider a ($1$-)form-valued $k$-form [$k>2$, the values $k=1$ and $%
k=2$ were previously analyzed \cite{LP15}] that `lives' in a $D$-dimensional
Minkowski spacetime. The coefficients of this ingredient constitute the
components of a tensor gauge field of degree $\left( k+1\right) $ that
transform under a reducible representation of the Lorentz group whose
irreducible decomposition involves two Young diagrams with one and
respectively two columns. This way geometrically put on the same foot the
tensor gauge fields of spin-$1$ [that transforms under irreducible
representations of Lorentz group pictured by one-column Young diagrams \cite%
{knaep1}, \cite{BB}] and the tensor gauge fields of spin-$2$ [that pertain
to the linear representations spaces of irreducible representations of the
Lorentz group encoded in Young diagrams with two columns \cite{BB}]. In view
of dynamically similarly behaviours for the `irreducible' components of our
basic object, we consider for this a PT-invariant, second-order Lagrangian
action that is labeled by two arbitrary real constants and that reduces [for
particular choices of the just mentioned real parameters] to the standard
Lagrangian actions for the Abelian $\left( k+1\right) $-form \cite{knaep1}
and that of a tensor gauge field with the mixed symmetry $\left( k,1\right) $
\cite{BB}. The benefits of such a unification come with the rich gauge
structure displayed by the Lagrangian theory aforementioned and mainly
consist in possible `exotic' consistent interactions that can be added among
one $\left( k+1\right) $-form and one tensor gauge field with the mixed
symmetry $\left( k,1\right) $.

This paper is organized into five sections as follows. In Section \ref%
{setting}, we start with a ($1$-)form-valued $k$-form and interpret it as a
collection of $k$-forms with a vector index. Then, we construct the most
general PT-invariant and second-order Lagrangian density that is invariant
under the standard gauge transformations of the just mentioned $k$-forms.
The local function depends on two arbitrary real constants [denoted by $a_{1}
$ and $a_{2}$] and, for some values of the real $a$-parameters, reduces to
standard Lagrangian densities corresponding to the Abelian $\left(
k+1\right) $-form \cite{knaep1} and that of a tensor gauge field with the
mixed symmetry $\left( k,1\right) $ \cite{BB}. At this stage, the natural
question appears: does the initial set of gauge transformations constitute a
generating one? It is the job of Section \ref{Dirac} to prove that the
answer to this question is mostly negative. Here, we perform the canonical
analysis \cite{DIRAC1}, \cite{DIRAC2}, \cite{1q} of the starting Lagrangian
theory. The procedure put into light a partition of the real parameters
plane $\left( a_{1},~a_{2}\right) $ made by seven components. For six among
the seven partition's components it is shown the generating set of gauge
transformations is \emph{richer} than the original gauge transformations,
including BF \cite{12} and/or conformal-like \cite{CONF}\ gauge
transformations. Moreover, in each of the seven situations is computed the
number of degrees of freedom and is investigated the presence of unphysical
degrees of freedom [ghost-modes]. It is proved the ghost-modes \emph{are
absent} only in two of the seven partition's components namely when the
Lagrangian density reduces to that of a Abelian $\left( k+1\right) $-form
and respectively to that of a tensor gauge field with the mixed symmetry $%
\left( k,1\right) $. These outputs generalize the previous results \cite%
{DESER}. In view of future investigations concerning the consistent
interactions that can be added among one $\left( k+1\right) $-form and one
tensor gauge field with the mixed symmetry $\left( k,1\right) $ in the
context of the considered ($1$-)form-valued $k$-form, the Section \ref{FO}
deals with the first-order formulations associated with the analyzed
second-order Lagrangian theory. Here, for each of the seven partition's
components, we generate the first-order Lagrangian formulation. These are
done with the price of adding auxiliary gauge/matter fields that make
possible the linearization. Section \ref{conclRGT} ends the paper with the
main conclusions.

\section{Setting the problem\label{setting}}

Our main ingredient is a ($1$-)form-valued $k$-form%
\begin{equation}
\mathbf{A}=\frac{1}{k!}A_{\mu _{1}\cdots \mu _{k}\parallel \alpha }\left(
\mathrm{d}x^{\mu _{1}}\cdots \mathrm{d}x^{\mu _{k}}\right) \otimes \mathrm{d}%
x^{\alpha },  \label{ingredient}
\end{equation}%
that `lives' in the $D$-dimensional Minkowski space [$D>k+1$] of `mostly
minus' signature [$\sigma _{\mu \nu }=\sigma ^{\mu \nu }=\mathrm{diag}\left(
+,-,\cdots ,-\right) $]. Its coordinates, $A_{\mu _{1}\cdots \mu
_{k}\parallel \alpha }$, are the components of a bosonic tensor gauge field
of degree $\left( k+1\right) $ that is antisymmetric in the first $k$
Lorentz indices
\begin{equation*}
A_{\mu _{1}\cdots \mu _{k}\parallel \alpha }=\frac{1}{k!}A_{\left[ \mu
_{1}\cdots \mu _{k}\right] \parallel \alpha }
\end{equation*}%
and with no symmetry in respect to the last, so one can interpret (\ref%
{ingredient}) in terms of a collection of $k$-forms%
\begin{equation*}
\left\{ \mathbf{A}_{\alpha }:\alpha =\overline{0,D-1}\right\} ,\quad \mathbf{%
A}_{\alpha }=\frac{1}{k!}A_{\mu _{1}\cdots \mu _{k}\parallel \alpha }\left(
\mathrm{d}x^{\mu _{1}}\cdots \mathrm{d}x^{\mu _{k}}\right) .
\end{equation*}%
Based on these, we are justified to postulate for the fields $A_{\mu
_{1}\cdots \mu _{k}\parallel \alpha }$ the gauge transformations%
\begin{equation}
\delta _{\epsilon }A_{\mu _{1}\cdots \mu _{k}\parallel \alpha }=\partial
_{\lbrack \mu _{1}}\epsilon _{\mu _{2}\cdots \mu _{k}]\parallel \alpha }.
\label{3m1}
\end{equation}%
In the above, the bosonic gauge parameters $\epsilon _{\mu _{1}\cdots \mu
_{k-1}\parallel \alpha }$ are completely antisymmetric in theirs first $%
\left( k-1\right) $ Lorentz indices%
\begin{equation*}
\epsilon _{\mu _{1}\cdots \mu _{k-1}\parallel \alpha }=\frac{1}{\left(
k-1\right) !}\epsilon _{\left[ \mu _{1}\cdots \mu _{k-1}\right] \parallel
\alpha }.
\end{equation*}%
The notation $[\mu \ldots \nu ]$ signifies full antisymmetry with respect to
the indices between brackets without normalization factors [i.e. the
independent terms appear only once and are not multiplied by overall
numerical factors]. In terms of the starting point (\ref{ingredient}), the
gauge transformations (\ref{3m1}) can be written as%
\begin{equation}
\delta _{\epsilon }\mathbf{A}=\mathrm{d}\mathbf{\epsilon },  \label{3m1a}
\end{equation}%
where
\begin{equation*}
\mathbf{\epsilon }=\frac{1}{\left( k-1\right) !}\epsilon _{\mu _{1}\cdots
\mu _{k-1}\parallel \alpha }\left( \mathrm{d}x^{\mu _{1}}\cdots \mathrm{d}%
x^{\mu _{k-1}}\right) \otimes \mathrm{d}x^{\alpha }
\end{equation*}%
is the gauge parameter vector-valued $\left( k-1\right) $-form. In the gauge
transformation (\ref{3m1a}) we used the notation $\mathrm{d}$ for the de
Rham differential in the exterior algebra $\bigwedge \mathbb{M}_{D}$. At
this stage, from the perspective of the linear representations of the
Lorentz group, the tensor gauge field $A_{\mu _{1}\cdots \mu _{k}\parallel
\alpha }$ pertains to the reducible representation space%
\begin{equation}
\begin{tabular}{c}
$\underbrace{%
\begin{tabular}{|l|}
\hline
$\mu _{1}$ \\ \hline
$\vdots $ \\ \hline
$\mu _{k}$ \\ \hline
\end{tabular}%
\otimes
\begin{tabular}{|l|}
\hline
$\alpha $ \\ \hline
\end{tabular}%
}$ \\
$A_{\mu _{1}\cdots \mu _{k}\parallel \alpha }$%
\end{tabular}%
\simeq
\begin{tabular}{c}
$\underbrace{%
\begin{tabular}{|l|}
\hline
$\mu _{1}$ \\ \hline
$\vdots $ \\ \hline
$\mu _{k}$ \\ \hline
$\alpha $ \\ \hline
\end{tabular}%
}$ \\
$B_{\mu _{1}\cdots \mu _{k}\alpha }$%
\end{tabular}%
\oplus
\begin{tabular}{c}
$\underbrace{%
\begin{tabular}{|l|l}
\hline
$\mu _{1}$ & \multicolumn{1}{|l|}{$\alpha $} \\ \hline
$\vdots $ &  \\ \cline{1-1}
$\mu _{k}$ &  \\ \cline{1-1}
\end{tabular}%
}$ \\
$t_{\mu _{1}\cdots \mu _{k}\mid \alpha }$%
\end{tabular}%
.  \label{3red}
\end{equation}

Now, we are interested in identifying the most general second-order
Lagrangian density that does not break the PT-invariance and is invariant
under the gauge transformations (\ref{3m1}) [or equivalently (\ref{3m1a})].
In view of this, based on the gauge transformations (\ref{3m1a}), one finds
the vector valued $\left( k+1\right) $-form%
\begin{eqnarray}
\mathbf{F} &\equiv &\mathrm{d}\mathbf{A}=\frac{1}{\left( k+1\right) !}%
\partial _{\lbrack \mu _{1}}A_{\mu _{2}\cdots \mu _{k+1}]\parallel \alpha
}\left( \mathrm{d}x^{\mu _{1}}\cdots \mathrm{d}x^{\mu _{k+1}}\right) \otimes
\mathrm{d}x^{\alpha }  \notag \\
&\equiv &\frac{1}{\left( k+1\right) !}F_{\mu _{1}\cdots \mu _{k+1}\parallel
\alpha }\left( \mathrm{d}x^{\mu _{1}}\cdots \mathrm{d}x^{\mu _{k+1}}\right)
\otimes \mathrm{d}x^{\alpha }  \label{3m1b}
\end{eqnarray}%
that is manifestly gauge-invariant under (\ref{3m1a}). The seeked Lagrangian
density, can be written in terms of the field-strength's coeffiecients as%
\begin{equation}
\mathcal{L}_{0}=\tfrac{1}{2\left( k+1\right) \left( k+1\right) !}\left( -%
\tfrac{\left( -\right) ^{k}}{k+1}\left( F_{\mu _{1}\cdots \mu
_{k+1}\parallel \alpha }\right) ^{2}+a_{1}F_{\mu _{1}\cdots \mu _{k}\beta
\parallel \alpha }F^{\mu _{1}\cdots \mu _{k}\alpha \parallel \beta
}+a_{2}\left( F^{\mu _{1}\cdots \mu _{k}}\right) ^{2}\right) ,  \label{3m2}
\end{equation}%
where $a_{1}$ and $a_{2}$ are arbitrary real constants. Moreover, by $F_{\mu
_{1}\cdots \mu _{k}}$ we denoted the trace of the field-strength, $F_{\mu
_{1}\cdots \mu _{k}}\equiv \sigma ^{\alpha \beta }F_{\mu _{1}\cdots \mu
_{k}\beta \parallel \alpha }$ and also we employed the notation%
\begin{equation*}
\left( U_{\Delta }\right) ^{2}\equiv \left( U^{\Delta }\right) ^{2}\equiv
U_{\Delta }U^{\Delta }
\end{equation*}%
for the Lorentz multi-index $\Delta =\mu _{1}\cdots \mu _{\delta }$
contractions.

In this point , we prove that the Lagrangian density (\ref{3m2}) can be used
to treat in a unitary manner two tensor gauge fields that transform under
irreducible representations of the Lorentz group, namely $\left( k+1\right) $%
-forms and the tensor gauge fields with the mixed symmetry $\left(
k,1\right) $. Accordingly with the isomorphysm in (\ref{3red}), we decompose
the gauge field $A_{\mu _{1}\cdots \mu _{k}\parallel \alpha }$ into its
'irreducible' components
\begin{equation}
A_{\mu _{1}\cdots \mu _{k}\parallel \alpha }\equiv B_{\mu _{1}\cdots \mu
_{k}\alpha }+t_{\mu _{1}\cdots \mu _{k}\mid \alpha }\equiv \left( \tfrac{1}{%
k+1}A_{\left[ \mu _{1}\cdots \mu _{k}\parallel \alpha \right] }\right)
+\left( A_{\mu _{1}\cdots \mu _{k}\parallel \alpha }-\tfrac{1}{k+1}A_{\left[
\mu _{1}\cdots \mu _{k}\parallel \alpha \right] }\right) .  \label{3m4}
\end{equation}%
Replacing this split into the definition of the field-strength (\ref{3m1b})
we get%
\begin{equation}
F_{\mu _{1}\cdots \mu _{k+1}\parallel \alpha }=H_{\mu _{1}\cdots \mu
_{k+1}\alpha }+\left( -\right) ^{k}\partial _{\alpha }B_{\mu _{1}\cdots \mu
_{k+1}}+\mathcal{F}_{\mu _{1}\cdots \mu _{k+1}\mid \alpha },  \label{3m5}
\end{equation}%
where we employed the notations%
\begin{equation}
H_{\mu _{1}\cdots \mu _{k+1}\alpha }\equiv \partial _{\lbrack \mu
_{1}}B_{\mu _{2}\cdots \mu _{k+1}\alpha ]},\quad \mathcal{F}_{\mu _{1}\cdots
\mu _{k+1}\mid \alpha }\equiv \partial _{\lbrack \mu _{1}}t_{\mu _{2}\cdots
\mu _{k+1}]\mid \alpha }  \label{3m6}
\end{equation}%
for the field-strengths corresponding to the 'irreducible' components $%
B_{\mu _{1}\cdots \mu _{k}\alpha }$ and $t_{\mu _{1}\cdots \mu _{k}\mid
\alpha }$. Based on the result (\ref{3m5}), by direct computations, we bring
the Lagrangian density (\ref{3m2})\ under the form%
\begin{eqnarray}
\mathcal{L}_{0} &=&-\tfrac{\left( -\right) ^{k}+\left( k^{2}+k+1\right)
a_{1}+a_{2}}{2\left( k+1\right) ^{2}\left( k+2\right) !}\left( H_{\mu
_{1}\cdots \mu _{k+2}}\right) ^{2}+\tfrac{a_{1}+a_{2}-\left( -\right) ^{k}}{%
2\left( k+1\right) ^{2}\left( k+1\right) !}\left( \partial _{\alpha }B_{\mu
_{1}\cdots \mu _{k+1}}\right) ^{2}  \notag \\
&&+\tfrac{a_{1}-\left( -\right) ^{k}}{2\left( k+1\right) ^{2}\left(
k+1\right) !}\left( \mathcal{F}_{\mu _{1}\cdots \mu _{k+1}\mid \alpha
}\right) ^{2}+\tfrac{a_{2}}{2\left( k+1\right) \left( k+1\right) !}\left(
\mathcal{F}_{\mu _{1}\cdots \mu _{k}}\right) ^{2}  \notag \\
&&+\left( -\right) ^{k}\tfrac{a_{1}+a_{2}-\left( -\right) ^{k}}{\left(
k+1\right) ^{2}\left( k+1\right) !}\mathcal{F}_{\mu _{1}\cdots \mu
_{k+1}\mid \alpha }\partial ^{\alpha }B^{\mu _{1}\cdots \mu _{k+1}}+\partial
_{\mu }j^{\mu }.  \label{3m7}
\end{eqnarray}%
Here $\mathcal{F}_{\mu _{1}\cdots \mu _{k}}$ is nothing but the trace of the
tensor $\mathcal{F}_{\mu _{1}\cdots \mu _{k+1}\mid \alpha }$ [$\mathcal{F}%
_{\mu _{1}\cdots \mu _{k}}\equiv \sigma ^{\alpha \beta }\mathcal{F}_{\mu
_{1}\cdots \mu _{k}\alpha \mid \beta }$]. Also, the components of the local
current in the left hand side of the density (\ref{3m5}) have the concrete
expressions%
\begin{eqnarray}
j^{\mu } &=&\tfrac{a_{2}}{2\left( k+1\right) \left( k+1\right) !}\left(
B^{\mu \mu _{1}\cdots \mu _{k}}\partial ^{\rho }B_{\rho \mu _{1}\cdots \mu
_{k}}-B_{\rho \mu _{1}\cdots \mu _{k}}\partial ^{\rho }B^{\mu \mu _{1}\cdots
\mu _{k}}\right.  \notag \\
&&\left. +2B^{\mu \mu _{1}\cdots \mu _{k}}\mathcal{F}_{\mu _{1}\cdots \mu
_{k}}-\tfrac{2\left( -\right) ^{k}}{k+1}B_{\mu _{1}\cdots \mu _{k+1}}%
\mathcal{F}^{\mu _{1}\cdots \mu _{k+1}\mid \mu }\right) .  \label{3m8}
\end{eqnarray}%
The Lagrangian action based on the local function (\ref{3m7})
\begin{eqnarray}
&&S_{0}^{\mathrm{L}}\left[ B_{\mu _{1}\cdots \mu _{k+1}},t_{\mu _{1}\cdots
\mu _{k}\mid \alpha }\right] =\int \mathrm{d}^{D}x\left[ \left( -\right) ^{k}%
\tfrac{a_{1}+a_{2}-\left( -\right) ^{k}}{\left( k+1\right) ^{2}\left(
k+1\right) !}\mathcal{F}_{\mu _{1}\cdots \mu _{k+1}\mid \alpha }\partial
^{\alpha }B^{\mu _{1}\cdots \mu _{k+1}}\right.  \notag \\
&&-\tfrac{\left( -\right) ^{k}k+\left( k^{2}+k+1\right) a_{1}+a_{2}}{2\left(
k+1\right) ^{2}\left( k+2\right) !}\left( H_{\mu _{1}\cdots \mu
_{k+2}}\right) ^{2}+\tfrac{a_{1}+a_{2}-\left( -\right) ^{k}}{2\left(
k+1\right) ^{2}\left( k+1\right) !}\left( \partial _{\alpha }B_{\mu
_{1}\cdots \mu _{k+1}}\right) \partial ^{\alpha }B^{\mu _{1}\cdots \mu
_{k+1}}  \notag \\
&&\left. +\tfrac{a_{1}-\left( -\right) ^{k}}{2\left( k+1\right) ^{2}\left(
k+1\right) !}\left( \mathcal{F}_{\mu _{1}\cdots \mu _{k+1}\mid \alpha
}\right) ^{2}+\tfrac{a_{2}}{2\left( k+1\right) \left( k+1\right) !}\left(
\mathcal{F}_{\mu _{1}\cdots \mu _{k}}\right) ^{2}\right]  \label{3m9}
\end{eqnarray}%
governs the dynamics of the tensor fields $t_{\mu _{1}\cdots \mu _{k}\mid
\alpha }$ and $B_{\mu _{1}\cdots \mu _{k}\alpha }$ that transform under
irreducible representations of the Lorentz group. In (\ref{3m9}) the
components $t_{\mu _{1}\cdots \mu _{k}\mid \alpha }$ and $B_{\mu _{1}\cdots
\mu _{k}\alpha }$ do not mix iff the constants $a_{1}$ and $a_{2}$ are
subjects to the algebraic equation%
\begin{equation}
a_{1}+a_{2}-\left( -\right) ^{k}=0\Leftrightarrow a_{2}=\left( -\right)
^{k}-a_{1}.  \label{3m10}
\end{equation}%
Indeed, by inserting the solution (\ref{3m10}) in the right hand side of (%
\ref{3m9}) we get%
\begin{equation}
S_{0}^{\mathrm{L}}\left[ B_{\mu _{1}\cdots \mu _{k+1}},t_{\mu _{1}\cdots \mu
_{k}\mid \alpha }\right] =\tfrac{\left( -\right) ^{k}+ka_{1}}{k+1}%
S_{0}^{\left( \mathrm{k+1}\right) }\left[ B_{\mu _{1}\cdots \mu _{k+1}}%
\right] +\tfrac{a_{1}-\left( -\right) ^{k}}{\left( k+1\right) ^{2}}%
S_{0}^{\left( \mathrm{k,1}\right) }\left[ t_{\mu _{1}\cdots \mu _{k}\mid
\alpha }\right] ,  \label{3m11}
\end{equation}%
where $S_{0}^{\left( \mathrm{k+1}\right) }\left[ B_{\mu _{1}\cdots \mu
_{k+1}}\right] $ and $S_{0}^{\left( \mathrm{k,1}\right) }\left[ t_{\mu
_{1}\cdots \mu _{k}\mid \alpha }\right] $ are the standard actions for $%
\left( k+1\right) $-form \cite{knaep1} and respectively for the massless
tensor gauge field with the mixed symmetry $\left( k,1\right) $ \cite{BB}.
Now, if we set in (\ref{3m11})
\begin{equation}
a_{1}=\left( -\right) ^{k},  \label{3m12}
\end{equation}%
the $\left( k+1\right) $-form $B_{\mu _{1}\cdots \mu _{k+1}}$ becomes a pure
gauge field. Also, tacking in (\ref{3m11})
\begin{equation}
a_{1}=-\frac{\left( -\right) ^{k}}{k},  \label{3m13}
\end{equation}%
the field $t_{\mu _{1}\cdots \mu _{k}\mid \alpha }$\ with the mixed symmetry
$\left( k,1\right) $ becomes a pure gauge one.

The above analysis allows us to conclude that some tensor gauge fields of
degree $\left( k+1\right) $ that transform under irreducible representations
of the Lorentz group can be treated in unified manner through the gauge
field $A_{\mu _{1}\cdots \mu _{k}\parallel \alpha }$ whose dynamics is
generated by the Lagrangian density (\ref{3m2}).

\section{Dirac analysis\label{Dirac}}

In this section we perform the canonical analysis \cite{DIRAC1}, \cite%
{DIRAC2}, \cite{1q} of the model with the Lagrangian density (\ref{3m2}). In
view of this, if we denote by $\pi _{\mu _{1}\cdots \mu _{k}\parallel \alpha
}$ the canonical momenta associated with the fields $A^{\mu _{1}\cdots \mu
_{k}\parallel \alpha }$, the definitions of the formers read as
\begin{eqnarray}
\pi _{\mu _{1}\cdots \mu _{k}\parallel \alpha } &\equiv &\frac{1}{k!}\frac{%
\partial \mathcal{L}_{0}}{\partial \dot{A}^{\left[ \mu _{1}\cdots \mu _{k}%
\right] \parallel \alpha }}  \notag \\
&=&\tfrac{\left( -\right) ^{k+1}}{\left( k+1\right) \left( k+1\right) !}%
\left( F_{0\mu _{1}\cdots \mu _{k}\parallel \alpha }-a_{1}F_{\alpha \left[
0\mu _{1}\cdots \mu _{k-1}\parallel \mu _{k}\right] }-a_{2}\sigma _{\alpha
\lbrack 0}F_{\mu _{1}\cdots \mu _{k}]}\right)   \label{c1}
\end{eqnarray}%
where by overdot we denoted the derivative in respect with the temporal
coordinate $x^{0}$. From the definitions in the above, we infer the primary
constraints%
\begin{equation}
G_{i_{1}\cdots i_{k-1}}^{(1)}\equiv \pi _{0i_{1}\cdots i_{k-1}\parallel
0}\approx 0,\quad G_{i_{1}\cdots i_{k-1}\parallel j}^{(1)}\equiv \pi
_{0i_{1}\cdots i_{k-1}\parallel j}\approx 0  \label{c2}
\end{equation}%
and also the relations%
\begin{eqnarray}
\pi _{i_{1}\cdots i_{k}\parallel 0} &=&\tfrac{1}{\left( k+1\right) \left(
k+1\right) !}\left[ \left( a_{1}+a_{2}-\left( -\right) ^{k}\right)
F_{0i_{1}\cdots i_{k}\parallel 0}+\left( -\right) ^{k}a_{2}F_{i_{1}\cdots
i_{k}}^{\prime }\right] ,  \label{c3} \\
\pi _{i_{1}\cdots i_{k}\parallel j} &=&\tfrac{\left( -\right) ^{k+1}}{\left(
k+1\right) \left( k+1\right) !}\left( F_{0i_{1}\cdots i_{k}\parallel
j}-a_{1}F_{j\left[ 0i_{1}\cdots i_{k-1}\parallel i_{k}\right] }-\left(
-\right) ^{k}a_{2}\sigma _{j[i_{1}}F_{i_{2}\cdots i_{k}]0}\right) .
\label{c4}
\end{eqnarray}%
In formula (\ref{c3}) we denoted by $F_{i_{1}\cdots i_{k}}^{\prime }$ the
spatial part of the field-strength's trace [$F_{i_{1}\cdots i_{k}}^{\prime
}\equiv \sigma ^{jl}F_{i_{1}\cdots i_{k}j\parallel l}$]. In the flow of the
analysis we will also use the 'irreducible' components of the primary
constraints (\ref{c2})%
\begin{eqnarray}
G_{i_{1}\cdots i_{k-1}\parallel j}^{(1)} &\equiv &\left( \tfrac{1}{k}\pi _{0%
\left[ i_{1}\cdots i_{k-1}\parallel j\right] }\right) +\left( \pi
_{0i_{1}\cdots i_{k-1}\parallel j}-\tfrac{1}{k}\pi _{0\left[ i_{1}\cdots
i_{k-1}\parallel j\right] }\right)   \notag \\
&\equiv &\left( \tfrac{1}{k}G_{\left[ i_{1}\cdots i_{k-1}\parallel j\right]
}^{(1)}\right) +\left( G_{i_{1}\cdots i_{k-1}\parallel j}^{(1)}-\tfrac{1}{k}%
G_{\left[ i_{1}\cdots i_{k-1}\parallel j\right] }^{(1)}\right) \equiv
G_{i_{1}\cdots i_{k}}^{(1)}+G_{i_{1}\cdots i_{k-1}\mid j}^{(1)}.  \label{c2d}
\end{eqnarray}%
Based on the equations (\ref{c4}), only by algebraic computations, we derive
\begin{eqnarray}
\pi _{i_{1}\cdots i_{k-1}}^{\prime } &\equiv &\sigma ^{jk}\pi _{i_{1}\cdots
i_{k-1}j\parallel k}=\tfrac{a_{1}+a_{2}\left( D-k\right) -\left( -\right)
^{k}}{\left( k+1\right) \left( k+1\right) !}F_{0i_{1}\cdots i_{k-1}},
\label{c5} \\
\pi _{\left[ i_{1}\cdots i_{k}\parallel i_{k+1}\right] } &=&\tfrac{\left(
-\right) ^{k}}{\left( k+1\right) !}\left[ a_{1}F_{i_{1}\cdots
i_{k+1}\parallel 0}-\left( -\right) ^{k}\tfrac{ka_{1}+\left( -\right) ^{k}}{%
\left( k+1\right) }F_{0\left[ i_{1}\cdots i_{k}\parallel i_{k+1}\right] }%
\right] .  \label{c6}
\end{eqnarray}%
The first step in the canonical analysis is achived by solving the equations
(\ref{c3})--(\ref{c4}) in respect with some of the generalized velocities.
In view of this, the results (\ref{c5})--(\ref{c6}) lead to seven distinct
situations [dictated by the factors that multiply the \textit{temporal}
\textit{components} of the field-strength in (\ref{c3})--(\ref{c4})], namely%
\begin{eqnarray}
a_{1} &=&\left( -\right) ^{k},\quad a_{2}=0;  \label{c7i} \\
a_{1} &=&-\tfrac{\left( -\right) ^{k}}{k},\quad a_{2}=\left( -\right) ^{k}%
\tfrac{k+1}{k};  \label{c7ii} \\
a_{1} &=&-\tfrac{\left( -\right) ^{k}}{k},\quad a_{2}=\left( -\right) ^{k}%
\tfrac{k+1}{k\left( D-k\right) };  \label{c7iii} \\
a_{1} &=&-\tfrac{\left( -\right) ^{k}}{k},\quad a_{2}\equiv a\in \mathbb{R}%
\backslash \left\{ \left( -\right) ^{k}\tfrac{k+1}{k},\left( -\right) ^{k}%
\tfrac{k+1}{k\left( D-k\right) }\right\} ;  \label{c7iv} \\
a_{1} &\equiv &\bar{a}\in
\mathbb{R}
\backslash \left\{ -\tfrac{\left( -\right) ^{k}}{k},\left( -\right)
^{k}\right\} ,\quad a_{2}=\tfrac{\left( -\right) ^{k}-\bar{a}}{D-k};
\label{c7v} \\
a_{1} &\equiv &\tilde{a}\in
\mathbb{R}
\backslash \left\{ -\tfrac{\left( -\right) ^{k}}{k},\left( -\right)
^{k}\right\} ,\quad a_{2}=\left( -\right) ^{k}-\tilde{a};  \label{c7vi} \\
a_{1}+a_{2} &\neq &\left( -\right) ^{k}\neq a_{1}+\left( D-k\right)
a_{2},\quad a_{1}\in
\mathbb{R}
\backslash \left\{ \left( -\right) ^{k}\right\} .  \label{c7vii}
\end{eqnarray}

In the remaining part of this section we will complete the canonical
analysis of the model in each of the seven situations delimited in the
above. This will include a careful analysis of the nature of independent
degrees of freedom [physical/ghost modes].

\subsection{Case I\label{I}}

In the situation (\ref{c7i}) the Lagrangian density (\ref{3m2}) becomes%
\begin{equation}
\mathcal{L}_{0}^{\left( I\right) }=\tfrac{\left( -\right) ^{k}}{2\left(
k+1\right) \left( k+1\right) !}\left( -\tfrac{1}{k+1}F_{\mu _{1}\cdots \mu
_{k+1}\parallel \alpha }F^{\mu _{1}\cdots \mu _{k+1}\parallel \alpha
}+F_{\mu _{1}\cdots \mu _{k}\beta \parallel \alpha }F^{\mu _{1}\cdots \mu
_{k}\alpha \parallel \beta }\right)  \label{ic1}
\end{equation}%
and the definitions of the canonical momenta (\ref{c1}) lead to the
independent primary constraints (\ref{c2}) and%
\begin{eqnarray}
\gamma _{i_{1}\cdots i_{k}}^{(1)} &\equiv &\pi _{i_{1}\cdots i_{k}\parallel
0}\approx 0,  \label{ic4} \\
\gamma _{i_{1}\cdots i_{k}\mid j}^{(1)} &\equiv &\pi _{i_{1}\cdots
i_{k}\parallel j}-\tfrac{1}{k+1}\pi _{\left[ i_{1}\cdots i_{k}\parallel j%
\right] }\approx 0.  \label{ic5}
\end{eqnarray}%
Solving now the equations (\ref{c1}) [corresponding to the choice (\ref{c7i}%
) of the real parameters $a_{1}$ and $a_{2}$] in respect with some of the
generalized velocities, we get the canonical Hamiltonian
\begin{eqnarray}
\mathcal{H}_{0}^{\left( I\right) } &=&-kA^{0i_{1}\cdots i_{k-1}\parallel
j}\partial ^{l}\pi _{li_{1}\cdots i_{k-1}\parallel j}-\tfrac{\left( -\right)
^{k}k!}{2}\pi _{i_{1}\cdots i_{k}\parallel j}\pi ^{\left[ i_{1}\cdots
i_{k}\parallel j\right] }  \notag \\
&&+\tfrac{\left( -\right) ^{k}}{\left( k+1\right) ^{2}}\pi ^{\left[
i_{1}\cdots i_{k}\parallel j\right] }F_{i_{1}\cdots i_{k}j\parallel 0}+%
\tfrac{\left( -\right) ^{k}}{2\left( k+1\right) \left( k+2\right) !}%
F_{i_{1}\cdots i_{k}\parallel j}F^{\left[ i_{1}\cdots i_{k}\parallel j\right]
}.  \label{ic6}
\end{eqnarray}

As the primary constraints (\ref{c2}) and (\ref{ic4})--(\ref{ic5}) depend
only on the momenta we gather their Abelian character so that the
consistency of the primary constraints reduces only to the computation of
the Poisson brackets between them and the canonical Hamiltonian. In view of
these, simple computations lead to
\begin{equation}
\left[ G_{i_{1}\cdots i_{k-1}}^{(1)},\mathcal{H}_{0}^{\left( I\right) }%
\right] =0=\left[ \gamma _{i_{1}\cdots i_{k}\mid j}^{(1)},\mathcal{H}%
_{0}^{\left( I\right) }\right]  \label{ic6a}
\end{equation}%
and
\begin{eqnarray}
\left[ \gamma _{i_{1}\cdots i_{k}}^{(1)},\mathcal{H}_{0}^{\left( I\right) }%
\right] &=&\tfrac{1}{k+1}\partial ^{j}\pi _{\left[ i_{1}\cdots
i_{k}\parallel j\right] }\equiv \gamma _{i_{1}\cdots i_{k}}^{(2)}\approx 0,
\label{ic6b} \\
\left[ G_{i_{1}\cdots i_{k-1}\parallel j}^{(1)},\mathcal{H}_{0}^{\left(
I\right) }\right] &=&\partial ^{i}\gamma _{ii_{1}\cdots i_{k}\mid j}^{\left(
1\right) }+\left( -\right) ^{k}\gamma _{i_{1}\cdots i_{k-1}j}^{(2)}\approx 0,
\label{ic6c}
\end{eqnarray}%
results that reveal the secondary constraints%
\begin{equation}
\gamma _{i_{1}\cdots i_{k}}^{(2)}\equiv \tfrac{1}{k+1}\partial ^{j}\pi _{%
\left[ i_{1}\cdots i_{k}\parallel j\right] }\approx 0.  \label{ic7}
\end{equation}%
Invoking again the dependence of the constraints (\ref{c2}), (\ref{ic4})--(%
\ref{ic5}) and (\ref{ic7}) only on the canonical momenta, we establish
theirs Abelianity. This remark, together with the Poisson brackets
\begin{equation}
\left[ \gamma _{i_{1}\cdots i_{k}}^{(2)},\mathcal{H}_{0}^{\left( I\right) }%
\right] =0  \label{ic7a}
\end{equation}%
allows us to conclude that the Dirac algorithm stops at this stage.

In order to count the independent degrees of freedom, we invoke the
first-class character of the constraints set (\ref{c2}), (\ref{ic4})--(\ref%
{ic5}) and (\ref{ic7}) supplemented with the off-shell reducibilities of
order $L=k$ of the secondary constraints (\ref{ic7})%
\begin{eqnarray*}
\left( Z_{j_{1}\cdots j_{k-1}}\right) ^{i_{1}\cdots i_{k}}\gamma
_{i_{1}\cdots i_{k}}^{(2)} &=&0 \\
\left( Z_{l_{1}\cdots l_{k-p-2}}\right) ^{j_{1}\cdots j_{k-p-1}}\left(
Z_{j_{1}\cdots j_{k-p-1}}\right) ^{i_{1}\cdots i_{k-p}} &=&0,\quad p=%
\overline{0,k-2}.
\end{eqnarray*}%
In the above we used the notations%
\begin{equation}
\left( Z_{j_{1}\cdots j_{k-p-1}}\right) ^{i_{1}\cdots i_{k-p}}=\partial
_{\left. {}\right. }^{[i_{1}}\delta _{j_{1}}^{i_{2}}\cdots \delta
_{j_{k-p-1}}^{i_{k-p}]},\quad p=\overline{0,k-1}.  \label{ic8}
\end{equation}%
The arguments that we have just given allow us to conclude that: the
canonical Hamiltonian (\ref{ic6}) is of the first-class [so this is the
classical observable that governs the time evolution] and the number of
independent degrees of freedom for the model under study is%
\begin{equation}
N_{DOF}^{\left( I\right) }=\left(
\begin{array}{c}
D-2 \\
k+1%
\end{array}%
\right) .  \label{ic10}
\end{equation}

Next, we analyze the nature [physical/unphysical] of the degrees of freedom (%
\ref{ic10}). In view of this, we firstly pass to the reduced phase-space [by
choosing of some apropriate canonical gauge conditions]. Then, we evaluate
the kinetic term of the first-class Hamiltonian (\ref{ic6}) restricted to
the reduced phase-space. If the kinetic term possesses definitness
[negatively or positively] then all degrees of freedom are physical.
Otherwise, ghost modes are present among the degrees of freedom.

In our case, a set of canonical gauge conditions consists in%
\begin{eqnarray}
\chi ^{(1)i_{1}\cdots i_{k-1}} &\equiv &A^{0i_{1}\cdots i_{k-1}\parallel
0}\approx 0,\quad \chi ^{(1)i_{1}\cdots i_{k-1}\parallel j}\equiv
A^{0i_{1}\cdots i_{k-1}\parallel j}\approx 0,  \label{ic10a} \\
\bar{\chi}^{(1)i_{1}\cdots i_{k}} &\equiv &A^{i_{1}\cdots i_{k}\parallel
0}\approx 0,\quad \bar{\chi}^{(1)i_{1}\cdots i_{k}\mid j}\equiv
A^{i_{1}\cdots i_{k}\parallel j}-\tfrac{1}{k+1}A^{\left[ i_{1}\cdots
i_{k}\parallel j\right] }\approx 0  \label{ic10b}
\end{eqnarray}%
and%
\begin{equation}
\bar{\chi}^{(2)i_{1}\cdots i_{k}}\equiv \tfrac{1}{k+1}\partial _{j}A^{\left[
i_{1}\cdots i_{k}\parallel j\right] }\approx 0  \label{ic10c}
\end{equation}%
Now, looking at the restriction
\begin{equation}
\mathcal{\bar{H}}_{0}^{\left( I\right) }\approx -\tfrac{\left( -\right)
^{k}k!}{2}\pi _{i_{1}\cdots i_{k}\parallel j}\pi ^{\left[ i_{1}\cdots
i_{k}\parallel j\right] }+\tfrac{\left( -\right) ^{k}}{2\left( k+1\right)
\left( k+2\right) !}F_{i_{1}\cdots i_{k}\parallel j}F^{\left[ i_{1}\cdots
i_{k}\parallel j\right] }  \label{ic10d}
\end{equation}%
of the first-class Hamiltonian on the reduced phase space, we conclude that
all degrees of freedom (\ref{ic10}) are physicall.

Finally, based on the Dirac's conjecture [according to which any first-class
constraint generates gauge transformations], if we pass again to the
Lagrangian formulation [via extended action], we derive for the functional
\begin{equation}
S_{0}^{\left( I\right) }\left[ A_{\mu _{1}\cdots \mu _{k}\parallel \alpha }%
\right] =\int \mathrm{d}^{D}x\mathcal{L}_{0}^{\left( I\right) }  \label{ic11}
\end{equation}%
the generating set of gauge transformations%
\begin{equation}
\delta _{\epsilon ,\xi }^{\left( I\right) }A_{\mu _{1}\cdots \mu
_{k}\parallel \alpha }=\partial _{\lbrack \mu _{1}}\epsilon _{\mu _{2}\cdots
\mu _{k}\alpha ]}+\xi _{\mu _{1}\cdots \mu _{k}\mid \alpha },  \label{ic12}
\end{equation}%
where the gauge parameters $\xi _{\mu _{1}\cdots \mu _{k}\mid \alpha }$ have
the mixed symmetry $\left( k,1\right) $%
\begin{equation}
\xi _{\mu _{1}\cdots \mu _{k}\mid \alpha }=\tfrac{1}{k!}\xi _{\left[ \mu
_{1}\cdots \mu _{k}\right] \mid \alpha },\quad \xi _{\left[ \mu _{1}\cdots
\mu _{k}\mid \alpha \right] }=0.  \label{ic13}
\end{equation}

We observe that the results (\ref{ic12}) imply that the 'irreducible'
component $t_{\mu _{1}\cdots \mu _{k}\mid \alpha }$ [with the concrete
expression given in (\ref{3m4})] is a pure gauge field%
\begin{equation*}
\delta _{\epsilon ,\xi }^{\left( I\right) }t_{\mu _{1}\cdots \mu _{k}\mid
\alpha }=\xi _{\mu _{1}\cdots \mu _{k}\mid \alpha }
\end{equation*}%
output that agrees with the discution in the end of the previous subsection.

\subsection{Case II\label{II}}

Now, we complete the canonical analysis of the model (\ref{3m2}) in the
second situation [the real parameters $a_{1}$ and $a_{2}$ take the values (%
\ref{c7ii})]. In this context, the Lagrangian density (\ref{3m2}) becomes%
\begin{equation}
\mathcal{L}_{0}^{\left( II\right) }=\tfrac{\left( -\right) ^{k+1}}{2\left(
k+1\right) \left( k+1\right) !}\left[ \tfrac{1}{k+1}\left( F_{\mu _{1}\cdots
\mu _{k+1}\parallel \alpha }\right) ^{2}+\tfrac{1}{k}F_{\mu _{1}\cdots \mu
_{k}\beta \parallel \alpha }F^{\mu _{1}\cdots \mu _{k}\alpha \parallel \beta
}-\tfrac{k+1}{k}\left( F_{\mu _{1}\cdots \mu _{k}}\right) ^{2}\right] .
\label{iic1}
\end{equation}%
Based on the choice (\ref{c7ii}), the definitions of the canonical momenta (%
\ref{c1}) lead to the independent primary constraints (\ref{c2}) and%
\begin{eqnarray}
\bar{\gamma}_{i_{1}\cdots i_{k}}^{(1)} &\equiv &\pi _{i_{1}\cdots
i_{k}\parallel 0}-\tfrac{1}{k\cdot \left( k+1\right) !}F_{i_{1}\cdots
i_{k}}^{\prime }\approx 0,  \label{iic2} \\
\bar{\gamma}_{i_{1}\cdots i_{k+1}}^{(1)} &\equiv &\pi _{\left[ i_{1}\cdots
i_{k}\parallel i_{k+1}\right] }+\tfrac{1}{k\cdot \left( k+1\right) !}%
F_{i_{1}\cdots i_{k+1}\parallel 0}\approx 0.  \label{iic3}
\end{eqnarray}%
Solving the equations (\ref{c1}) in respect with some of the generalized
velocities, we derive the canonical Hamiltonian density [well defined only
on the primary constraint surface]%
\begin{eqnarray}
\mathcal{H}_{0}^{\left( II\right) } &=&-kA^{0i_{1}\cdots i_{k-1}\parallel
\mu }\left( \partial ^{l}\pi _{li_{1}\cdots i_{k-1}\parallel \mu }\right) +%
\tfrac{\left( -\right) ^{k}}{2\left( k+1\right) ^{2}\left( k+1\right) !}%
\left( F_{i_{1}\cdots i_{k+1}\parallel \mu }\right) ^{2}  \notag \\
&&+\tfrac{\left( -\right) ^{k}}{2k\left( k+1\right) \left( k+1\right) !}%
F_{i_{1}\cdots i_{k}j\parallel l}F^{i_{1}\cdots i_{k}l\parallel j}-\tfrac{%
\left( -\right) ^{k}}{2k\left( k+1\right) !}\left( F_{i_{1}\cdots
i_{k}}^{\prime }\right) ^{2}  \notag \\
&&-\tfrac{\left( -\right) ^{k}k\left( k+1\right) !}{2}\pi _{i_{1}\cdots
i_{k}\parallel j}\pi ^{i_{1}\cdots i_{k}\parallel j}-\tfrac{\left( -\right)
^{k}}{2\left( k+1\right) }\pi _{i_{1}\cdots i_{k}\parallel j}F^{i_{1}\cdots
i_{k}j\parallel 0}  \notag \\
&&+\tfrac{\left( -\right) ^{k}k^{2}\left( k+1\right) !}{2\left( D-k-1\right)
}\pi _{i_{1}\cdots i_{k-1}}^{\prime }\pi ^{\prime i_{1}\cdots i_{k-1}}.
\label{iic4}
\end{eqnarray}

The next step --- consistency of the primary constraints is solved in two
stages. Initially, by direct computation one infers the Abelian character of
the set of primary constraints consisting in (\ref{c2}) and (\ref{iic2})--(%
\ref{iic3}). This allow us to conclude that the consistency of the primary
constraints reduces only to the calculations between canonical Hamiltonian
and primary constraints. By direct computations one obtains%
\begin{eqnarray}
\left[ G_{i_{1}\cdots i_{k-1}}^{(1)},H_{0}^{\left( II\right) }\right]
&=&\partial ^{j}\pi _{ji_{1}\cdots i_{k-1}\parallel 0}\equiv G_{i_{1}\cdots
i_{k-1}}^{(2)}\approx 0,  \label{iic5a} \\
\left[ G_{i_{1}\cdots i_{k-1}\parallel j}^{(1)},H_{0}^{\left( II\right) }%
\right] &=&\partial ^{l}\pi _{li_{1}\cdots i_{k-1}\parallel j}\equiv
G_{i_{1}\cdots i_{k-1}\parallel j}^{(2)}\approx 0,  \label{iic5b} \\
\left[ \bar{\gamma}_{i_{1}\cdots i_{k}}^{(1)},H_{0}^{\left( II\right) }%
\right] &=&\tfrac{2k+1}{2\left( k+1\right) }\partial ^{j}\bar{\gamma}%
_{i_{1}\cdots i_{k}j}^{(1)}-G_{[i_{1}\cdots i_{k-1}\parallel
i_{k}]}^{(2)}\approx 0,  \label{iic5c} \\
\left[ \bar{\gamma}_{i_{1}\cdots i_{k+1}}^{(1)},H_{0}^{\left( II\right) }%
\right] &=&0.  \label{iic5d}
\end{eqnarray}%
The results (\ref{iic5a})--(\ref{iic5d}) display the secondary constraints%
\begin{equation}
G_{i_{1}\cdots i_{k-1}}^{(2)}\approx 0,\quad G_{i_{1}\cdots i_{k-1}\parallel
j}^{(2)}\approx 0,  \label{iic6}
\end{equation}%
that together with (\ref{c2}) and (\ref{iic2})--(\ref{iic3}) constitute an
Abelian set of constraints. These outputs, supplemented with the Poisson
brackets%
\begin{equation}
\left[ G_{i_{1}\cdots i_{k-1}}^{(2)},H_{0}^{\left( II\right) }\right] =0=%
\left[ G_{i_{1}\cdots i_{k-1}\parallel j}^{(2)},H_{0}^{\left( II\right) }%
\right]  \label{iic6a}
\end{equation}%
allow us to conclude that the Dirac algoritm stops at this stage.

In the next, in order to count the degrees of freedom, we use: the
first-class constraints (\ref{c2}), (\ref{iic2})--(\ref{iic3}) and (\ref%
{iic6}), the irreducible character of the constraints (\ref{c2}), (\ref{iic2}%
)--(\ref{iic3}) and the $L=k-1$ reducibilities of the secondary constraints (%
\ref{iic6})%
\begin{eqnarray}
\left( Z_{j_{1}\cdots j_{k-2}}\right) ^{i_{1}\cdots i_{k-1}}G_{i_{1}\cdots
i_{k-1}}^{(2)} &=&0,  \label{iic6ax} \\
\left( Z_{l_{1}\cdots l_{k-p-3}}\right) ^{j_{1}\cdots j_{k-p-2}}\left(
Z_{j_{1}\cdots j_{k-p-2}}\right) ^{i_{1}\cdots i_{k-p-1}} &=&0,\quad p=%
\overline{0,k-3},  \label{iic6ay} \\
\left( Z_{j_{1}\cdots j_{k-2}}\right) ^{i_{1}\cdots i_{k-1}\Vert
i}G_{i_{1}\cdots i_{k-1}\parallel i}^{(2)} &=&0,  \label{iic6az} \\
\left( Z_{l_{1}\cdots l_{k-p-3}\Vert l}\right) ^{j_{1}\cdots i_{k-p-2}\Vert
j}\left( Z_{j_{1}\cdots j_{k-p-2}\Vert j}\right) ^{i_{1}\cdots
i_{k-p-1}\Vert i} &=&0,\quad p=\overline{0,k-3}.  \label{iic6at}
\end{eqnarray}%
In the above, we used the notations%
\begin{eqnarray}
\left( Z_{j_{1}\cdots j_{k-p-2}}\right) ^{i_{1}\cdots i_{k-1-p}} &=&\partial
_{\left. {}\right. }^{[i_{1}}\delta _{j_{1}}^{i_{2}}\cdots \delta
_{j_{k-p-1}}^{i_{k-p}]},\quad p=\overline{0,k-2},  \label{iic7a} \\
\left( Z_{j_{1}\cdots j_{k-p-2}\Vert j}\right) ^{i_{1}\cdots i_{k-1-p}\Vert
i} &=&\delta _{j}^{i}\left( Z_{j_{1}\cdots j_{k-p-2}}\right) ^{i_{1}\cdots
i_{k-1-p}},\quad p=\overline{0,k-2}.  \label{iic7b}
\end{eqnarray}%
Putting these together we get the number of independent degrees of freedom
for the model under study
\begin{equation}
N_{DOF}^{\left( II\right) }=D\left(
\begin{array}{c}
D-2 \\
k%
\end{array}%
\right) -\left(
\begin{array}{c}
D \\
k+1%
\end{array}%
\right) .  \label{iic8}
\end{equation}

As in the previous situation, we are interested about the 'nature' of the
degrees o freedom. In order to do this, firstly we chose the set of the
canonical gauge conditions corresponding to the first-class constraints (\ref%
{c2}), (\ref{iic2})--(\ref{iic3}) and (\ref{iic6}) consisting in (\ref{ic10a}%
) and%
\begin{eqnarray}
\bar{\chi}^{\left( 1\right) i_{1}\cdots i_{k}} &\equiv &A^{i_{1}\cdots
i_{k}\parallel 0}\approx 0,  \label{iic9a} \\
\bar{\chi}^{\left( 1\right) i_{1}\cdots i_{k+1}} &\equiv &A^{\left[
i_{1}\cdots i_{k}\parallel i_{k+1}\right] }\approx 0,  \label{iic9b} \\
\bar{\chi}_{i_{1}\cdots i_{k-1}}^{\left( 1\right) } &\equiv &\pi
_{i_{1}\cdots i_{k-1}}^{\prime }\approx 0,  \label{iic9c} \\
\bar{\chi}^{\left( 2\right) i_{1}\cdots i_{k-1}} &\equiv &\partial
_{i}A^{ii_{1}\cdots i_{k-1}\parallel 0}\approx 0,  \label{iic9d} \\
\bar{\chi}^{\left( 2\right) i_{1}\cdots i_{k-1}\parallel j} &\equiv
&\partial _{i}A^{ii_{1}\cdots i_{k-1}\parallel j}\approx 0.  \label{iic9e}
\end{eqnarray}%
Evaluating now the restriction of the first-class Hamiltonian (\ref{iic4})
to the reduced phase space we get%
\begin{eqnarray}
\mathcal{\bar{H}}_{0}^{\left( II\right) } &\approx &-\tfrac{\left( -\right)
^{k}k\left( k+1\right) !}{2}\pi _{i_{1}\cdots i_{k}\parallel \mu }\pi
^{i_{1}\cdots i_{k}\parallel \mu }+\tfrac{\left( -\right) ^{k}}{2\left(
k+1\right) ^{2}\left( k+1\right) !}\left( F_{i_{1}\cdots i_{k+1}\parallel
j}\right) ^{2}  \notag \\
&&+\tfrac{\left( -\right) ^{k}}{2k\left( k+1\right) \left( k+1\right) !}%
F_{i_{1}\cdots i_{k}j\parallel l}F^{i_{1}\cdots i_{k}l\parallel j},
\label{iic9}
\end{eqnarray}
we conclude that also in this case ghost modes do not appear.

Finally, if we pass again to the Lagrangian formulation [via extended
action], we derive for the functional
\begin{equation}
S_{0}^{\left( II\right) }\left[ A_{\mu _{1}\cdots \mu _{k}\parallel \alpha }%
\right] =\int \mathrm{d}^{D}x\mathcal{L}_{0}^{\left( II\right) }
\label{iic10}
\end{equation}%
the generating set of gauge transformations%
\begin{equation}
\delta _{\epsilon ,\xi }^{\left( II\right) }A_{\mu _{1}\cdots \mu
_{k}\parallel \alpha }=\epsilon _{\mu _{1}\cdots \mu _{k}\alpha }+\partial
_{\lbrack \mu _{1}}\epsilon _{\mu _{2}\cdots \mu _{k}]\alpha }-\left(
-\right) ^{k}k\partial _{\alpha }\epsilon _{\mu _{1}\cdots \mu
_{k}}+\partial _{\lbrack \mu _{1}}\xi _{\mu _{2}\cdots \mu _{k}]\mid \alpha
},  \label{iic11}
\end{equation}%
where the gauge parameters $\epsilon _{\mu _{1}\cdots \mu _{k+1}}$ and $%
\epsilon _{\mu _{1}\cdots \mu _{k}}$ are completely antisymmetric%
\begin{equation}
\epsilon _{\mu _{1}\cdots \mu _{k+1}}=\tfrac{1}{\left( k+1\right) !}\epsilon
_{\lbrack \mu _{1}\cdots \mu _{k+1}]},\quad \epsilon _{\mu _{1}\cdots \mu
_{k}}=\tfrac{1}{k!}\epsilon _{\lbrack \mu _{1}\cdots \mu _{k}]}
\label{iic12}
\end{equation}%
while $\xi _{\mu _{1}\cdots \mu _{k-1}\mid \alpha }$ display the mixed
symmetry $\left( k-1,1\right) $%
\begin{equation}
\xi _{\mu _{1}\cdots \mu _{k-1}\mid \alpha }=\tfrac{1}{\left( k-1\right) !}%
\xi _{\lbrack \mu _{1}\cdots \mu _{k-1}]\mid \alpha },\quad \xi _{\lbrack
\mu _{1}\cdots \mu _{k-1}\mid \alpha ]}=0.  \label{iic13}
\end{equation}%
From (\ref{iic11}) we infer that the irreducible component $B_{\mu
_{1}\cdots \mu _{k+1}}$ is a pure gauge field%
\begin{equation}
\delta _{\epsilon ,\xi }^{\left( II\right) }B_{\mu _{1}\cdots \mu
_{k+1}}=\epsilon _{\mu _{1}\cdots \mu _{k+1}}.  \label{iic14}
\end{equation}

\subsection{Case III\label{III}}

For the choice (\ref{c7iii}) of the parameters $a_{1}$ and $a_{2}$ the
Lagrangian density (\ref{3m2}) takes the form%
\begin{equation}
\mathcal{L}_{0}^{\left( III\right) }=\tfrac{\left( -\right) ^{k+1}}{2\left(
k+1\right) \left( k+1\right) !}\left[ \tfrac{1}{k+1}\left( F_{\mu _{1}\cdots
\mu _{k+1}\parallel \alpha }\right) ^{2}+\tfrac{1}{k}F_{\mu _{1}\cdots \mu
_{k}\beta \parallel \alpha }F^{\mu _{1}\cdots \mu _{k}\alpha \parallel \beta
}-\tfrac{k+1}{k\left( D-k\right) }\left( F_{\mu _{1}\cdots \mu _{k}}\right)
^{2}\right] .  \label{iiic1}
\end{equation}%
Replacing (\ref{c7iii}) into the definitions (\ref{c1}) one infers the
primary constraints (\ref{c2}) and
\begin{eqnarray}
\gamma _{i_{1}\cdots i_{k-1}}^{(1)} &\equiv &\pi _{i_{1}\cdots
i_{k-1}}^{\prime }\approx 0,  \label{iiic2} \\
\bar{\gamma}_{i_{1}\cdots i_{k+1}}^{(1)} &\equiv &\pi _{\lbrack i_{1}\cdots
i_{k}\parallel i_{k+1}]}+\tfrac{1}{k\cdot \left( k+1\right) !}F_{i_{1}\cdots
i_{k+1}\parallel 0}\approx 0.  \label{iiic3}
\end{eqnarray}%
Moreover, solving the corresponding equations (\ref{c1}) in respect with
some of the generalized velocities, we derive the canonical Hamiltonian
density [well-defined only on the primary constraints surface]
\begin{eqnarray}
\mathcal{H}_{0}^{\left( III\right) } &=&-kA^{0i_{1}\cdots i_{k-1}\parallel
\mu }\left( \partial ^{l}\pi _{li_{1}\cdots i_{k-1}\parallel \mu }\right) +%
\tfrac{\left( -\right) ^{k}}{2\left( k+1\right) ^{2}\left( k+1\right) !}%
\left( F_{i_{1}\cdots i_{k+1}\parallel \mu }\right) ^{2}  \notag \\
&&-\left( -\right) ^{k}\tfrac{k\left( k+1\right) !}{2}\left( \tfrac{D-k}{%
D-k-1}\left( \pi _{i_{1}\cdots i_{k}\parallel 0}\right) ^{2}+\left( \pi
_{i_{1}\cdots i_{k}\parallel j}\right) ^{2}\right)  \notag \\
&&-\tfrac{\left( -\right) ^{k}}{2k\left( D-k-1\right) \left( k+1\right) !}%
\left( F_{i_{1}\cdots i_{k}}^{\prime }\right) ^{2}+\tfrac{\left( -\right)
^{k}}{2k\left( k+1\right) \left( k+1\right) !}F^{i_{1}\cdots i_{k}j\parallel
l}F_{i_{1}\cdots i_{k}l\parallel j}  \notag \\
&&+\tfrac{\left( -\right) ^{k}}{D-k-1}\pi _{i_{1}\cdots i_{k}\parallel
0}F^{\prime i_{1}\cdots i_{k}}-\tfrac{\left( -\right) ^{k}}{2\left(
k+1\right) }\pi _{i_{1}\cdots i_{k}\parallel j}F^{i_{1}\cdots
i_{k}j\parallel 0}.  \label{iiic4}
\end{eqnarray}%
As in the previous two situations, the primary constraints (\ref{c2}) and (%
\ref{iiic2})--(\ref{iiic3}) are Abelian so their consistency reduces to the
computation of the Poisson brackets between them and the canonical
Hamiltonian (\ref{iiic4}). By direct calculations we infer%
\begin{eqnarray}
\left[ G_{i_{1}\cdots i_{k-1}}^{(1)},H_{0}^{\left( III\right) }\right]
&=&\partial ^{j}\pi _{ji_{1}\cdots i_{k-1}\parallel 0}\equiv G_{i_{1}\cdots
i_{k-1}}^{(2)}\approx 0,  \label{iiic5a} \\
\left[ G_{i_{1}\cdots i_{k-1}\parallel j}^{(1)},H_{0}^{\left( III\right) }%
\right] &=&\partial ^{l}\pi _{li_{1}\cdots i_{k-1}\parallel j}\equiv
G_{i_{1}\cdots i_{k-1}\parallel j}^{(2)}\approx 0,  \label{iiic5b} \\
\left[ \gamma _{i_{1}\cdots i_{k-1}}^{(1)},H_{0}^{\left( III\right) }\right]
&=&\left( -\right) ^{k}G_{i_{1}\cdots i_{k-1}}^{(2)}\approx 0,
\label{iiic5c} \\
\left[ \bar{\gamma}_{i_{1}\cdots i_{k+1}}^{(1)},H_{0}^{\left( III\right) }%
\right] &=&-\left( -\right) ^{k}\left( \partial _{\lbrack i_{1}}\pi
_{i_{2}\cdots i_{k+1}]\parallel 0}-\tfrac{\left( -\right) ^{k}}{k\left(
k+1\right) !}\partial ^{j}F_{i_{1}\cdots i_{k+1}\parallel j}\right)  \notag
\\
&\equiv &-\left( -\right) ^{k}\bar{\gamma}_{i_{1}\cdots
i_{k+1}}^{(2)}\approx 0.  \label{iiic5d}
\end{eqnarray}%
The results (\ref{iiic5a})--(\ref{iiic5d}) put into evidence the secondary
constraints%
\begin{equation}
G_{i_{1}\cdots i_{k-1}}^{(2)}\approx 0,\quad G_{i_{1}\cdots i_{k-1}\parallel
j}^{(2)}\approx 0,\quad \bar{\gamma}_{i_{1}\cdots i_{k+1}}^{(2)}\approx 0.
\label{iiic6}
\end{equation}

Direct computations show that the Poisson brackets among the constraints (%
\ref{c2}), (\ref{iiic2})--(\ref{iiic3}) and (\ref{iiic6}) are vanishing so
that the requirement of conservation in time for the secondary constraints (%
\ref{iiic6}) reduces, as in the previous situation, to the computation of
the Poisson brackets between (\ref{iiic6}) and canonical Hamiltonian (\ref%
{iiic4})%
\begin{equation}
\left[ G_{i_{1}\cdots i_{k-1}}^{(2)},H_{0}^{\left( III\right) }\right] =0=%
\left[ G_{i_{1}\cdots i_{k-1}\parallel j}^{(2)},H_{0}^{\left( III\right) }%
\right]  \label{iiic7a}
\end{equation}%
\begin{equation}
\left[ \bar{\gamma}_{i_{1}\cdots i_{k+1}}^{(2)},H_{0}^{\left( III\right) }%
\right] =\tfrac{\left( -\right) ^{k+1}}{k+1}\partial _{\lbrack
i_{1}}^{\left. {}\right. }G_{i_{2}\cdots i_{k}\Vert i_{k+1}]}^{(2)}+\tfrac{%
2k+1}{2\left( k+1\right) }\partial ^{j}\partial _{\lbrack i_{1}}^{\left.
{}\right. }\bar{\gamma}_{i_{2}\cdots i_{k+1}]j}^{(1)}\approx 0
\label{iiic7b}
\end{equation}

The previous analysis allow us to conclude that Dirac algorithm stops at
this stage and, moreover, the canonical Hamiltonian (\ref{iiic4}) coincides
with the first-class Hamiltonian of the system.

In order to count the independent degrees of freedom for the model under
study, we investigate the reducibilities of the first-class constraints set (%
\ref{c2}), (\ref{iiic2})--(\ref{iiic3}) and (\ref{iiic6}). The concrete
expressions of the analyzed constraints evidence that: i) the constraints $%
\bar{\gamma}_{i_{1}\cdots i_{k+1}}^{(2)}\approx 0$ are off-shell reducible
of order $\left( D-k-2\right) $ with the reducibility functions%
\begin{equation}
\left( Z_{j_{1}\cdots j_{k+p+1}}\right) ^{i_{1}\cdots i_{k+p}}=\partial
_{\lbrack j_{1}}^{\left. {}\right. }\delta _{j_{2}}^{i_{1}}\cdots \delta
_{j_{k+p+1}]}^{i_{k+p}},\quad k=\overline{1,D-k-2},  \label{iiic8}
\end{equation}%
ii) the constraints (\ref{iic6}) are $L=k-1$ off-shell reducible with the
reducibility functions given in (\ref{iic7a})--(\ref{iic7b}), iii) the
constraints $\gamma _{i_{1}\cdots i_{k-1}}^{(1)}\approx 0$ and $%
G_{i_{1}\cdots i_{k-1}\parallel j}^{(2)}\approx 0$ are off-shell first order
reducible%
\begin{equation}
\left( \partial _{\left. {}\right. }^{[i_{1}}\delta _{j_{1}}^{i_{2}}\cdots
\delta _{j_{k-2}}^{i_{k-1}]}\right) \gamma _{i_{1}\cdots
i_{k-1}}^{(1)}+\left( -\delta _{j_{1}}^{[i_{1}}\cdots \delta
_{j_{k-2}}^{i_{k-2}}\sigma _{\left. {}\right. }^{i_{k-1}]j}\right)
G_{i_{1}\cdots i_{k-1}\parallel j}^{(2)}=0,  \label{iiic9}
\end{equation}%
and iv) the constraints (\ref{c2}) and (\ref{iiic3}) are irreducible. The
previous reducibilities of the first-class constraints implies that the
number of independent degrees of freedom is%
\begin{equation}
N_{DOF}^{\left( III\right) }=D\left(
\begin{array}{c}
D-2 \\
k%
\end{array}%
\right) -\left(
\begin{array}{c}
D \\
k+1%
\end{array}%
\right) +\left(
\begin{array}{c}
D-2 \\
k-3%
\end{array}%
\right) .  \label{iiic10}
\end{equation}

At this stage we are interested if all the (\ref{iiic10}) independent
degrees of freedom are physical. In order to aswer to this question we
choose a set of canonical gauge conditions consisting in (\ref{ic10a}), (\ref%
{iic9b}), (\ref{iic9d})--(\ref{iic9e}) and
\begin{eqnarray}
\tilde{\chi}^{\left( 1\right) i_{1}\cdots i_{k-1}} &\equiv &A^{\prime
i_{1}\cdots i_{k-1}}\approx 0,  \label{iiic10a} \\
\bar{\chi}^{\left( 2\right) i_{1}\cdots i_{k+1}} &\equiv &\partial ^{\lbrack
i_{1}}A^{i_{2}\cdots i_{k+1}]\Vert 0}\approx 0.  \label{iiic10b}
\end{eqnarray}%
Evaluating now the first-class Hamiltonian (\ref{iiic4}) in the presence of
the canonical gauge conditions we obtain%
\begin{eqnarray}
\mathcal{\bar{H}}_{0}^{\left( III\right) } &\approx &\tfrac{k\left(
k+1\right) !}{2}\left( \pi _{i_{1}\cdots i_{k}\parallel j}^{2}-\tfrac{D-k}{%
D-k-1}\pi _{i_{1}\cdots i_{k}\parallel 0}^{2}\right)  \notag \\
&&+\tfrac{\left( -\right) ^{k}}{2\left( k+1\right) ^{2}\left( k+1\right) !}%
\left( F_{i_{1}\cdots i_{k+1}\parallel j}\right) ^{2}+\tfrac{\left( -\right)
^{k}}{2k\left( k+1\right) \left( k+1\right) !}F^{i_{1}\cdots i_{k}j\parallel
l}F_{i_{1}\cdots i_{k}l\parallel j}.  \label{iiic10c}
\end{eqnarray}%
Analyzing now the expression (\ref{iiic10c}) we observe that the modes $\pi
_{i_{1}\cdots i_{k}\parallel 0}$ comes with negative contributions in the
kinetic term so we conclude that ghost modes are present in this situation.

Finally, the returning to the Lagrangian formulation [via the extended
action] furnishes for the functional
\begin{equation}
S_{0}^{\left( III\right) }\left[ A_{\mu _{1}\cdots \mu _{k}\parallel \alpha }%
\right] =\int \mathrm{d}^{D}x\mathcal{L}_{0}^{\left( III\right) }
\label{iiic11}
\end{equation}%
the generating set of gauge transformations%
\begin{equation}
\delta _{\epsilon ,\xi }^{\left( III\right) }A_{\mu _{1}\cdots \mu
_{k}\parallel \alpha }=\sigma _{\alpha \lbrack \mu _{1}}\epsilon _{\mu
_{2}\cdots \mu _{k}]}+\partial ^{\sigma }\epsilon _{\mu _{1}\cdots \mu
_{k}\alpha \sigma }+\partial _{\lbrack \mu _{1}}\epsilon _{\mu _{2}\cdots
\mu _{k}]\alpha }+\partial _{\lbrack \mu _{1}}\xi _{\mu _{2}\cdots \mu
_{k}]\mid \alpha },  \label{iiic12}
\end{equation}%
where the gauge parameters of $\epsilon $-type are completely antisymmetric
and those of $\xi $-type possess the mixed symmetry $\left( k-1,1\right) $.
It is easy to see, that in the present situation, the gauge transformations
of the irreducible components become%
\begin{eqnarray}
\delta _{\epsilon ,\xi }^{\left( III\right) }t_{\mu _{1}\cdots \mu _{k}\mid
\alpha } &=&\sigma _{\alpha \lbrack \mu _{1}}\epsilon _{\mu _{2}\cdots \mu
_{k}]}+\tfrac{1}{k+1}\left( \partial _{\lbrack \mu _{1}}\epsilon _{\mu
_{2}\cdots \mu _{k}]\alpha }-\left( -\right) ^{k}k\partial _{\alpha
}\epsilon _{\mu _{1}\cdots \mu _{k}}\right)  \notag \\
&&+\partial _{\lbrack \mu _{1}}\xi _{\mu _{2}\cdots \mu _{k}]\mid \alpha },
\label{iiic13a} \\
\delta _{\epsilon ,\xi }^{\left( III\right) }B_{\mu _{1}\cdots \mu
_{k}\alpha } &=&\partial ^{\sigma }\epsilon _{\mu _{1}\cdots \mu _{k}\alpha
\sigma }+\tfrac{k}{k+1}\partial _{\lbrack \mu _{1}}\epsilon _{\mu _{2}\cdots
\mu _{k}\alpha ]}.  \label{iiic13b}
\end{eqnarray}

It is woth noticing the presence in the generating set (\ref{iiic12}) of
some conformal-like [first term in the right-hand side of (\ref{iiic12})]
and BF-type [second component in the right-hand side of (\ref{iiic12})]\
gauge transformations.

\subsection{Case IV\label{IV}}

From the dynamical point of view, this situation is quite similar to the
previous one as we shall see in the following. Making the choice (\ref{c7iv}%
) in the local function (\ref{3m2}), the Lagrangian density becomes%
\begin{equation}
\mathcal{L}_{0}^{\left( IV\right) }\equiv \tfrac{\left( -\right) ^{k+1}}{%
2\left( k+1\right) \left( k+1\right) !}\left[ \tfrac{1}{k+1}\left( F_{\mu
_{1}\cdots \mu _{k+1}\parallel \alpha }\right) ^{2}+\tfrac{1}{k}F_{\mu
_{1}\cdots \mu _{k}\beta \parallel \alpha }F^{\mu _{1}\cdots \mu _{k}\alpha
\parallel \beta }-\left( -\right) ^{k}a\left( F_{\mu _{1}\cdots \mu
_{k}}\right) ^{2}\right] ,  \label{ivc1}
\end{equation}%
where $a$ is an arbitrary real constant with the range given in (\ref{c7iv}).

In this context, the definitions (\ref{c1}) of the canonical momenta lead to
the\ Abelian primary constraints (\ref{c2}) and (\ref{iiic3}) and also
produce the canonical Hamiltonian density [well defined only on the primary
constraint surface]%
\begin{eqnarray}
\mathcal{H}_{0}^{\left( IV\right) } &=&-kA^{0i_{1}\cdots i_{k-1}\parallel
\mu }\left( \partial ^{l}\pi _{li_{1}\cdots i_{k-1}\parallel \mu }\right) +%
\tfrac{\left( -\right) ^{k}}{2\left( k+1\right) ^{2}\left( k+1\right) !}%
\left( F_{i_{1}\cdots i_{k+1}\parallel \mu }\right) ^{2}  \notag \\
&&-\left( -\right) ^{k}\tfrac{k\left( k+1\right) !}{2}\left( \tfrac{k+1}{%
k+1-\left( -\right) ^{k}ka}\left( \pi _{i_{1}\cdots i_{k}\parallel 0}\right)
^{2}+\left( \pi _{i_{1}\cdots i_{k}\parallel j}\right) ^{2}\right)  \notag \\
&&-\tfrac{ak^{3}\left( k+1\right) !}{2\left[ k+1-\left( -\right)
^{k}ak\left( D-k\right) \right] }\left( \pi _{i_{1}\cdots i_{k-1}}^{\prime
}\right) ^{2}+\tfrac{ka}{k+1-\left( -\right) ^{k}ka}\pi _{i_{1}\cdots
i_{k}\parallel 0}F^{\prime i_{1}\cdots i_{k}}  \notag \\
&&-\tfrac{a}{2\left( k+1-\left( -\right) ^{k}ka\right) \left( k+1\right) !}%
\left( F_{i_{1}\cdots i_{k}}^{\prime }\right) ^{2}+\tfrac{\left( -\right)
^{k}}{2k\left( k+1\right) \left( k+1\right) !}F^{i_{1}\cdots i_{k}j\parallel
l}F_{i_{1}\cdots i_{k}l\parallel j}  \notag \\
&&-\tfrac{\left( -\right) ^{k}}{2\left( k+1\right) }\pi _{i_{1}\cdots
i_{k}\parallel j}F^{i_{1}\cdots i_{k}j\parallel 0}.  \label{ivc2}
\end{eqnarray}

The second step in the Dirac analysis --- consistency of the primary
constraints --- involves the computations between the canonical Hamiltonian
and the primary constraints. Due to the fact that the primary constraints
are Abelian, their consistency requirement produces the secondary
constraints (\ref{iiic6}) as
\begin{equation}
\left[ G_{i_{1}\cdots i_{k-1}}^{(1)},H_{0}^{\left( IV\right) }\right] \equiv
G_{i_{1}\cdots i_{k-1}}^{(2)},\quad \left[ G_{i_{1}\cdots i_{k-1}\parallel
j}^{(1)},H_{0}^{\left( IV\right) }\right] \equiv G_{i_{1}\cdots
i_{k-1}\parallel j}^{(2)},\quad \left[ \bar{\gamma}_{i_{1}\cdots
i_{k+1}}^{(1)},H_{0}^{\left( IV\right) }\right] \equiv -\left( -\right) ^{k}%
\bar{\gamma}_{i_{1}\cdots i_{k+1}}^{(2)}.  \label{ivc3}
\end{equation}

Concerning the consistency of the secondary constraints (\ref{iiic6}) this
does not imply new constraints because firstly, the constraints (\ref{c2}), (%
\ref{iiic3}) and (\ref{iiic6}) are Abelian and secondly, the Poisson
brackets between the canonical Hamiltonian (\ref{ivc2}) and secondary
constraints (\ref{iiic6}) reads as in (\ref{iiic7a})--(\ref{iiic7b}).

The previous results allow us to state that the Dirac algorithm stops at
this level, and, moreover, (\ref{ivc2}) is nothing but the first-class
Hamiltonian.

In order to count the independent degrees of freedom, we use the
reducibilities of the first-class constraints (\ref{c2}), (\ref{iiic3}) and (%
\ref{iiic6}) established in the previous situation. More precisely, the
constraints: (\ref{c2}) and (\ref{iiic3}) are irreducible; $G_{i_{1}\cdots
i_{k-1}}^{(2)}\approx 0$ and $G_{i_{1}\cdots i_{k-1}\parallel
j}^{(2)}\approx 0$ are off-shell reducible of order $L=k-1$ with the
reducibility functions given in (\ref{iic7a})--(\ref{iic7b}) and $\bar{\gamma%
}_{i_{1}\cdots i_{k+1}}^{(2)}\approx 0$ are off-shell reducible of order $%
\left( D-k-2\right) $ with the reducibility functions expressed in (\ref%
{iiic8}). In view of these, the number of independent degrees of freedom in
the present situation reads as%
\begin{equation}
N_{DOF}^{(IV)}=\left( D-1\right) \left(
\begin{array}{c}
D-2 \\
k%
\end{array}%
\right) -\left(
\begin{array}{c}
D-1 \\
k+1%
\end{array}%
\right) .  \label{ivc5}
\end{equation}

As in the preceding situation we are interested whether all the degrees of
freedom are physical. In order to answer the this question, we evaluate the
first-class Hamiltonian (\ref{ivc2}) on the reduced phase-space. To do so,
we choose the set of canonical gauge-fixing conditions consisting in (\ref%
{ic10a}), (\ref{iic9b}), (\ref{iic9d}), (\ref{iic9e}) and (\ref{iiic10b})
and the time-evolution generator corresponding to (\ref{ivc2}) takes the form%
\begin{eqnarray}
\mathcal{\bar{H}}_{0}^{\left( IV\right) } &\approx &-\tfrac{k\left(
k+1\right) \left( k+1\right) !}{2\left( k+1-\left( -\right) ^{k}ka\right) }%
\pi _{i_{1}\cdots i_{k}\parallel 0}^{2}+\tfrac{k\left( k+1\right) !}{2}\Pi
_{i_{1}\cdots i_{k}\parallel j}^{2}  \notag \\
&&-\tfrac{a^{2}k^{3}\left( D-k\right) \left( k+1\right) !}{8\left(
k+1-\left( -\right) ^{k}ak\left( D-k\right) \right) ^{2}}\pi _{i_{1}\cdots
i_{k-1}}^{\prime 2}  \notag \\
&&-\tfrac{a}{2\left( k+1-\left( -\right) ^{k}ka\right) \left( k+1\right) !}%
\left( F_{i_{1}\cdots i_{k}}^{\prime }\right) ^{2}  \notag \\
&&+\tfrac{\left( -\right) ^{k}}{2\left( k+1\right) ^{2}\left( k+1\right) !}%
F^{i_{1}\cdots i_{k+1}\parallel j}\left( F_{i_{1}\cdots i_{k+1}\parallel j}+%
\tfrac{\left( -\right) ^{k}}{k}F_{j[i_{1}\cdots i_{k}\parallel
i_{k+1}]}\right)  \label{ivc5a}
\end{eqnarray}%
where we employed the notations%
\begin{equation*}
\Pi _{i_{1}\cdots i_{k}\parallel j}\equiv \pi _{i_{1}\cdots i_{k}\parallel
j}+\tfrac{\left( -\right) ^{k}k}{2\left( k+1-\left( -\right) ^{k}ak\left(
D-k\right) \right) }\pi _{\lbrack i_{1}\cdots i_{k-1}}^{\prime }\sigma
_{i_{k}]j}^{\left. {}\right. }.
\end{equation*}%
Based on the expression (\ref{ivc5a}) we conclude that the unphysical
degrees of freedom [ghost modes] are still present.

Invoking again the Dirac's conjecture, we derive for the functional
\begin{equation}
S_{0}^{\left( IV\right) }\left[ A_{\mu _{1}\cdots \mu _{k}\parallel \alpha }%
\right] =\int \mathrm{d}^{D}x\mathcal{L}_{0}^{\left( IV\right) }
\label{ivc6}
\end{equation}%
the generating set of gauge transformations%
\begin{equation}
\delta _{\epsilon ,\xi }^{\left( IV\right) }A_{\mu _{1}\cdots \mu
_{k}\parallel \alpha }=\partial ^{\sigma }\epsilon _{\mu _{1}\cdots \mu
_{k}\alpha \sigma }+\partial _{\lbrack \mu _{1}}\epsilon _{\mu _{2}\cdots
\mu _{k}]\alpha }+\partial _{\lbrack \mu _{1}}\xi _{\mu _{2}\cdots \mu
_{k}]\mid \alpha },  \label{ivc7}
\end{equation}%
where the gauge parameters of $\epsilon $ and $\xi $-type have the
symmetries specified in the previous situation.

It is remarkable the presence in the generating set (\ref{ivc7}) of a
BF-type [first term in the right-hand side of (\ref{ivc7})]\ gauge component.

\subsection{Case V\label{V}}

Now, the constants that parametrize (\ref{3m2}) are taken as in (\ref{c7v}).
With this choice, the Lagrangian density (\ref{3m2}) becomes
\begin{equation}
\mathcal{L}_{0}^{\left( V\right) }=\tfrac{\left( -\right) ^{k+1}}{2\left(
k+1\right) \left( k+1\right) !}\left[ \tfrac{1}{k+1}\left( F_{\mu _{1}\cdots
\mu _{k+1}\parallel \alpha }\right) ^{2}-\left( -\right) ^{k}\bar{a}F_{\mu
_{1}\cdots \mu _{k}\beta \parallel \alpha }F^{\mu _{1}\cdots \mu _{k}\alpha
\parallel \beta }+\tfrac{\left( -\right) ^{k}\bar{a}-1}{D-k}\left( F_{\mu
_{1}\cdots \mu _{k}}\right) ^{2}\right] ,  \label{vc1}
\end{equation}%
where the range of the real constant $\bar{a}$ is given in (\ref{c7v}). In
this context, the definitions of the canonical momenta (\ref{c1}) furnish
the primary constraints (\ref{c2}) and (\ref{iiic2}) and also produce the
canonical Hamiltonian density
\begin{eqnarray}
\mathcal{H}_{0}^{\left( V\right) } &=&-kA^{0i_{1}\cdots i_{k-1}\parallel \mu
}\left( \partial ^{l}\pi _{li_{1}\cdots i_{k-1}\parallel \mu }\right) +%
\tfrac{\left( -\right) ^{k}}{2\left( k+1\right) ^{2}\left( k+1\right) !}%
\left( F_{i_{1}\cdots i_{k+1}\parallel \mu }\right) ^{2}  \notag \\
&&+\tfrac{\left( k+1\right) \left( k+1\right) !}{2\left( \bar{a}-\left(
-\right) ^{k}\right) }\left( \tfrac{D-k}{D-k-1}\left( \pi _{i_{1}\cdots
i_{k}\parallel 0}\right) ^{2}+\left( \pi _{i_{1}\cdots i_{k}\parallel
j}\right) ^{2}\right.  \notag \\
&&\left. -\tfrac{\bar{a}}{k\bar{a}+\left( -\right) ^{k}}\pi _{i_{1}\cdots
i_{k}\parallel j}\pi ^{\left[ i_{1}\cdots i_{k}\parallel j\right] }\right) +%
\tfrac{\left( -\right) ^{k}\bar{a}}{k\bar{a}+\left( -\right) ^{k}}\pi
_{i_{1}\cdots i_{k}\parallel j}F^{i_{1}\cdots i_{k}j\parallel 0}  \notag \\
&&+\tfrac{\left( -\right) ^{k}}{D-k-1}\pi _{i_{1}\cdots i_{k}\parallel
0}F^{\prime i_{1}\cdots i_{k}}-\tfrac{\bar{a}^{2}}{2\left( k+1\right) \left(
k+1\right) !\left( k\bar{a}+\left( -\right) ^{k}\right) }\left(
F_{i_{1}\cdots i_{k+1}\parallel 0}\right) ^{2}  \notag \\
&&+\tfrac{1}{2\left( k+1\right) \left( k+1\right) !}\left( \tfrac{\bar{a}%
-\left( -\right) ^{k}}{D-k-1}\left( F_{i_{1}\cdots i_{k}}^{\prime }\right)
^{2}-\bar{a}F^{i_{1}\cdots i_{k}j\parallel l}F_{i_{1}\cdots i_{k}l\parallel
j}\right) .  \label{vc2}
\end{eqnarray}

Due to the Abelian character of the primary constraints (\ref{c2}) and (\ref%
{iiic2}), the second step in the Dirac analysis reduces to the computation
of the Poisson brackets between the canonical Hamiltonian (\ref{vc2}) and
the primary constraints (\ref{c2}) and (\ref{iiic2}). Direct calculations
display%
\begin{eqnarray}
\left[ G_{i_{1}\cdots i_{k-1}}^{(1)},H_{0}^{\left( V\right) }\right]
&=&G_{i_{1}\cdots i_{k-1}}^{(2)},  \label{vc3a} \\
\left[ G_{i_{1}\cdots i_{k-1}\parallel j}^{(1)},H_{0}^{\left( V\right) }%
\right] &=&G_{i_{1}\cdots i_{k-1}\parallel j}^{(2)},  \label{vc3b} \\
\left[ \gamma _{i_{1}\cdots i_{k-1}}^{(1)},H_{0}^{\left( V\right) }\right]
&=&\left( -\right) ^{k}G_{i_{1}\cdots i_{k-1}}^{(2)},  \label{vc3c}
\end{eqnarray}%
where the functions that appear in the right-hand side of (\ref{vc3a})--(\ref%
{vc3c}) have the concrete expressions given in (\ref{iiic5a}) and (\ref%
{iiic5c}).

Due to the fact that the primary and secondary constraints depend only on
the canonical momenta we conclude these are Abelian. Moreover, the Dirac
algorithm stops at this level as the consistency requirements of the
secondary constraints $G_{i_{1}\cdots i_{k-1}}^{(2)}\approx 0$ and $%
G_{i_{1}\cdots i_{k-1}\parallel j}^{(2)}\approx 0$ no longer produce
tertiary constraints
\begin{equation}
\left[ G_{i_{1}\cdots i_{k-1}}^{(2)},H_{0}^{\left( V\right) }\right] =0=%
\left[ G_{i_{1}\cdots i_{k-1}\parallel j}^{(2)},H_{0}^{\left( V\right) }%
\right] .  \label{vc4}
\end{equation}

At this stage we infer that the canonical Hamiltonian (\ref{vc2}) is of the
first-class and also we are able to count the number of independent degrees
of freedom. In view of this, we invoke: i) the irreducible character of the
first-class constraints (\ref{c2}), ii) the first-order reducibilities (\ref%
{iiic9}) of the constraints (\ref{iiic2}) and $G_{i_{1}\cdots
i_{k-1}\parallel j}^{(2)}$, iii) the $L=k-1$ off-shell reducibilities of the
constraints $G_{i_{1}\cdots i_{k-1}}^{(2)}\approx 0$ and $G_{i_{1}\cdots
i_{k-1}\parallel j}^{(2)}\approx 0$ [the reducibility functions are given in
(\ref{iic7a})--(\ref{iic7b})]. Based on these arguments, the number of
independent degrees of freedom is%
\begin{equation}
N_{DOF}^{\left( V\right) }=D\left(
\begin{array}{c}
D-2 \\
k%
\end{array}%
\right) -\left(
\begin{array}{c}
D-1 \\
k-1%
\end{array}%
\right) +\left(
\begin{array}{c}
D-1 \\
k-2%
\end{array}%
\right) .  \label{vc5}
\end{equation}

As in the previous cases we ask for the nature of independent degrees of
freedom. In order to answer to this question we select the canonical gauge
conditions (\ref{ic10a}), (\ref{iic9d}), (\ref{iic9e}) and (\ref{iiic10a})
and evaluate the restriction of the first-class Hamiltonian on the reduced
phase-space%
\begin{eqnarray}
\mathcal{\bar{H}}_{0}^{\left( V\right) } &\approx &\left( -\right) ^{k}%
\tfrac{\left( k+1\right) \left( k+1\right) !}{2\left( \bar{a}-\left(
-\right) ^{k}\right) }\left( \tfrac{D-k}{D-k-1}\Pi _{i_{1}\cdots
i_{k}\parallel 0}^{2}-\pi _{i_{1}\cdots i_{k}\parallel j}^{2}\right.  \notag
\\
&&\left. +\tfrac{1}{\left( k\bar{a}+\left( -\right) ^{k}\right) \left(
k+1\right) }\Pi _{i_{1}\cdots i_{k+1}}^{2}\right) +\tfrac{1}{2\left(
k+1\right) \left( k+1\right) !}\tfrac{\bar{a}-\left( -\right) ^{k}}{D-k}%
\left( F_{i_{1}\cdots i_{k}}^{\prime }\right) ^{2}  \notag \\
&&+\tfrac{1+\left( -\right) ^{k}\left( k-1\right) \bar{a}-k\bar{a}^{2}}{%
2\left( k+1\right) ^{2}\left( k+1\right) !\left( k\bar{a}+\left( -\right)
^{k}\right) }\left( F_{i_{1}\cdots i_{k+1}\parallel 0}\right) ^{2}  \notag \\
&&+\tfrac{\left( -\right) ^{k}}{2\left( k+1\right) ^{2}\left( k+1\right) !}%
F^{i_{1}\cdots i_{k+1}\parallel \mu }\left( F_{i_{1}\cdots i_{k+1}\parallel
j}-\bar{a}F_{j\left[ i_{1}\cdots i_{k}\parallel i_{k+1}\right] }\right) .
\label{vc5a}
\end{eqnarray}%
In the above we employed the notations%
\begin{eqnarray*}
\Pi _{i_{1}\cdots i_{k}\parallel 0} &=&\pi _{i_{1}\cdots i_{k}\parallel 0}+%
\tfrac{\left( -\right) ^{k}\left( \bar{a}-\left( -\right) ^{k}\right) }{%
\left( D-k\right) \left( k+1\right) \left( k+1\right) !}F_{i_{1}\cdots
i_{k}}^{\prime }, \\
\Pi _{i_{1}\cdots i_{k+1}} &=&\pi _{\left[ i_{1}\cdots i_{k}\parallel i_{k+1}%
\right] }-\tfrac{\bar{a}-\left( -\right) ^{k}}{\left( k+1\right) \left(
k+1\right) !}F_{i_{1}\cdots i_{k+1}\parallel 0}
\end{eqnarray*}%
Analyzing now the kinetic term of the generator of time-evolution (\ref{vc5a}%
) we conclude that also in this case the ghost modes are present.

Using the same method as in the previous subsections, one can deduce for the
functional
\begin{equation}
S_{0}^{\left( V\right) }\left[ A_{\mu _{1}\cdots \mu _{k}\parallel \alpha }%
\right] =\int \mathrm{d}^{D}x\mathcal{L}_{0}^{\left( V\right) }  \label{vc6}
\end{equation}%
the generating set of gauge transformations%
\begin{equation}
\delta _{\epsilon ,\xi }^{\left( V\right) }A_{\mu _{1}\cdots \mu
_{k}\parallel \alpha }=\sigma _{\alpha \lbrack \mu _{1}}\epsilon _{\mu
_{2}\cdots \mu _{k}]}+\partial _{\lbrack \mu _{1}}\epsilon _{\mu _{2}\cdots
\mu _{k}]\alpha }+\partial _{\lbrack \mu _{1}}\xi _{\mu _{2}\cdots \mu
_{k}]\mid \alpha }.  \label{vc7}
\end{equation}

It is woth noticing the presence in the generating set (\ref{vc7}) of a
conformal-like [first term in the right-hand side of (\ref{vc7})] gauge
transformation.

\subsection{Case VI\label{VI}}

Here we finish the canonical analysis when the real constants $a_{1}$ and $%
a_{2}$ have the values (\ref{c7vi}). For this setting, the Lagrangian
density (\ref{3m2}) becomes%
\begin{equation}
\mathcal{L}_{0}^{\left( VI\right) }=\tfrac{1}{2\left( k+1\right) \left(
k+1\right) !}\left[ \tfrac{\left( -\right) ^{k+1}}{k+1}\left( F_{\mu
_{1}\cdots \mu _{k+1}\parallel \alpha }\right) ^{2}+\tilde{a}F_{\mu
_{1}\cdots \mu _{k}\beta \parallel \alpha }F^{\mu _{1}\cdots \mu _{k}\alpha
\parallel \beta }+\left( \left( -\right) ^{k}-\tilde{a}\right) \left( F_{\mu
_{1}\cdots \mu _{k}}\right) ^{2}\right] .  \label{vic1}
\end{equation}

With the choice (\ref{c7vi}), the definitions (\ref{c1}) lead to the primary
constraints (\ref{c2}) and
\begin{equation}
\tilde{\gamma}_{i_{1}\cdots i_{k}}^{(1)}\equiv \pi _{i_{1}\cdots
i_{k}\parallel 0}+\frac{\left( -\right) ^{k}\tilde{a}-1}{\left( k+1\right)
\left( k+1\right) !}F_{i_{1}\cdots i_{k}}^{\prime }\approx 0.  \label{vic2}
\end{equation}%
Performing the Legendre transformation of (\ref{vic1}) in respect with some
of the generalized velocities [those that can be solved in the definitions
of the canonical momenta (\ref{c1})], we get the canonical Hamiltonian
density [well defined only on the primary constraint surface]%
\begin{eqnarray}
\mathcal{H}_{0}^{\left( VI\right) } &=&-kA^{0i_{1}\cdots i_{k-1}\parallel
\mu }\left( \partial ^{l}\pi _{li_{1}\cdots i_{k-1}\parallel \mu }\right) +%
\tfrac{\left( -\right) ^{k}}{2\left( k+1\right) ^{2}\left( k+1\right) !}%
\left( F_{i_{1}\cdots i_{k+1}\parallel j}\right) ^{2}  \notag \\
&&+\tfrac{\left( k+1\right) \left( k+1\right) !}{2\left( \tilde{a}-\left(
-\right) ^{k}\right) }\left( \left( \pi _{i_{1}\cdots i_{k}\parallel
j}\right) ^{2}-\tfrac{k}{D-k-1}\left( \pi _{i_{1}\cdots i_{k-1}}^{\prime
}\right) ^{2}\right.  \notag \\
&&\left. -\tfrac{\tilde{a}}{k\tilde{a}+\left( -\right) ^{k}}\pi
_{i_{1}\cdots i_{k}\parallel j}\pi ^{\left[ i_{1}\cdots i_{k}\parallel j%
\right] }\right) +\tfrac{\left( -\right) ^{k}\tilde{a}}{k\tilde{a}+\left(
-\right) ^{k}}\pi _{i_{1}\cdots i_{k}\parallel j}F^{i_{1}\cdots
i_{k}j\parallel 0}  \notag \\
&&+\tfrac{\tilde{a}-\left( -\right) ^{k}}{2\left( k+1\right) \left(
k+1\right) !}\left( F_{i_{1}\cdots i_{k}}^{\prime }\right) ^{2}+\tfrac{%
1+\left( -\right) ^{k}k\tilde{a}-\left( k+1\right) \tilde{a}^{2}}{2\left(
k+1\right) ^{2}\left( k+1\right) !\left( k\tilde{a}+\left( -\right)
^{k}\right) }\left( F_{i_{1}\cdots i_{k+1}\parallel 0}\right) ^{2}  \notag \\
&&-\tfrac{\tilde{a}}{2\left( k+1\right) \left( k+1\right) !}F^{i_{1}\cdots
i_{k}j\parallel l}F_{i_{1}\cdots i_{k}l\parallel j}.  \label{vic3}
\end{eqnarray}

The second step in the canonical analysis --- the time-preservation of the
primary constraints --- reduces to the computation of the Poisson brackets
and primary constraints. This is due to the fact that the primary
constraints (\ref{c2}) and (\ref{vic2}) are Abelian. In this light, simple
calculations lead to%
\begin{equation}
\left[ G_{i_{1}\cdots i_{k-1}}^{(1)},H_{0}^{\left( VI\right) }\right]
=G_{i_{1}\cdots i_{k-1}}^{(2)},\quad \left[ G_{i_{1}\cdots i_{k-1}\parallel
j}^{(1)},H_{0}^{\left( VI\right) }\right] =G_{i_{1}\cdots i_{k-1}\parallel
j}^{(2)},\quad \left[ \tilde{\gamma}_{i_{1}\cdots i_{k}}^{(1)},H_{0}^{\left(
VI\right) }\right] =-\tilde{\gamma}_{i_{1}\cdots i_{k}}^{(2)},  \label{vic4}
\end{equation}%
where the functions in the right-hand sides are given in formulas (\ref%
{iiic5a}), (\ref{iiic5b}) and
\begin{equation}
\tilde{\gamma}_{i_{1}\cdots i_{k}}^{(2)}\equiv -\partial ^{j}\left( \pi
_{i_{1}\cdots i_{k}\parallel j}+\tfrac{\left( -\right) ^{k}-\tilde{a}}{%
\left( k+1\right) \left( k+1\right) !}F_{ji_{1}\cdots i_{k}\parallel
0}\right) .  \label{vic5}
\end{equation}%
These results derived in the above allow to display the secondary
constraints possessed by the model under study%
\begin{equation}
G_{i_{1}\cdots i_{k-1}}^{(2)}\approx 0,\quad G_{i_{1}\cdots i_{k-1}\parallel
j}^{(2)}\approx 0,\quad \tilde{\gamma}_{i_{1}\cdots i_{k}}^{(2)}\approx 0.
\label{vic6}
\end{equation}

At this stage we investigate the consistency of the secondary constraints.
By direct computation we deduce that the all the constraints (\ref{c2}), (%
\ref{vic2}) and (\ref{vic6}) are are Abelian and, moreover, the Poisson
brackets hold%
\begin{equation}
\left[ G_{i_{1}\cdots i_{k-1}}^{(2)},H_{0}^{\left( VI\right) }\right]
=0,\quad \left[ G_{i_{1}\cdots i_{k-1}\parallel j}^{(2)},H_{0}^{\left(
VI\right) }\right] =0=\left[ \tilde{\gamma}_{i_{1}\cdots
i_{k}}^{(2)},H_{0}^{\left( VI\right) }\right] .  \label{vic7}
\end{equation}%
Based on these arguments we conclude that the model under study possesses no
tertiary constraints and, in addition, the canonical Hamiltonian (\ref{vic6}%
) coincides with the first-class Hamiltonian.

In the light of the counting of independent degrees of freedom, we make use
of the argumentation: i) the first-class constraints (\ref{c2}) and (\ref%
{vic2}) are irreducible and ii) the two subsets in the secondary constraints
(\ref{vic6}) are off-shell reducible of order $L=k-1$ with the relations of
reducibility given in (\ref{iic6ax})--(\ref{iic6at}) and the last two ones
are off-shell reducible of order $L=k$ with the reducibility relations%
\begin{eqnarray}
\left( \tilde{Z}_{j_{1}\cdots j_{k-1}}\right) ^{i_{1}\cdots i_{k-1}\Vert
i}G_{i_{1}\cdots i_{k-1}\parallel i}^{(2)}+\left( \tilde{Z}_{j_{1}\cdots
j_{k-1}}\right) ^{i_{1}\cdots i_{k}}\tilde{\gamma}_{i_{1}\cdots i_{k}}^{(2)}
&=&0,  \label{vic8a} \\
\left( \tilde{Z}_{l_{1}\cdots l_{k-2}}\right) ^{j_{1}\cdots j_{k-1}}\left(
\tilde{Z}_{j_{1}\cdots j_{k-1}}\right) ^{i_{1}\cdots i_{k-1}\Vert i} &=&0,
\label{vic8b} \\
\left( \tilde{Z}_{l_{1}\cdots l_{k-2}}\right) ^{j_{1}\cdots j_{k-1}}\left(
\tilde{Z}_{j_{1}\cdots j_{k-1}}\right) ^{i_{1}\cdots i_{k}} &=&0,
\label{vic8c} \\
\left( \tilde{Z}_{l_{1}\cdots l_{k-p-3}}\right) ^{j_{1}\cdots
j_{k-p-2}}\left( \tilde{Z}_{j_{1}\cdots j_{k-p-2}}\right) ^{i_{1}\cdots
i_{k-p-1}} &=&0\quad p=\overline{0,k-3}.  \label{vic8d}
\end{eqnarray}%
In the above, we used the notations%
\begin{eqnarray}
\left( \tilde{Z}_{j_{1}\cdots j_{k-1}}\right) ^{i_{1}\cdots i_{k-1}\Vert i}
&=&k\partial _{\left. {}\right. }^{[i}\delta _{j_{1}}^{i_{1}}\cdots \delta
_{j_{k-1}}^{i_{k-1}]},\quad \left( \tilde{Z}_{j_{1}\cdots j_{k-1}}\right)
^{i_{1}\cdots i_{k}}=\partial _{\left. {}\right. }^{[i_{1}}\delta
_{j_{1}}^{i_{1}}\cdots \delta _{j_{k-1}}^{i_{k}]}  \label{vic8e} \\
\left( \tilde{Z}_{j_{1}\cdots j_{k-p-1}}\right) ^{i_{1}\cdots i_{k-p}}
&=&\partial _{\left. {}\right. }^{[i_{1}}\delta _{j_{1}}^{i_{2}}\cdots
\delta _{j_{k-p-1}}^{i_{k-p}]},\quad p=\overline{1,k-1}.  \label{vic8f}
\end{eqnarray}%
Putting together the previous results we determine the number of independent
degrees of freedom
\begin{equation}
N_{DOF}^{\left( VI\right) }=\left( D-1\right) \left(
\begin{array}{c}
D-2 \\
k%
\end{array}%
\right) -\left(
\begin{array}{c}
D-1 \\
k%
\end{array}%
\right) .  \label{vic9}
\end{equation}

As in the situations previously analyzed, at this level we are interested if
there are ghost modes among the independent degrees o freedom (\ref{vic9}).
In view of this, we firstly choose the canonical gauge conditions (\ref%
{ic10a}), (\ref{iic9a}), (\ref{iic9c})--(\ref{iic9e}) and
\begin{equation}
\tilde{\chi}^{\left( 2\right) i_{1}\cdots i_{k}}\equiv \partial
_{j}A^{i_{1}\cdots i_{k}\Vert j}\approx 0  \label{vic10}
\end{equation}%
and then we investigate the kinetic term in the restriciton of the
first-class Hamiltonian (\ref{vic3}) on the reduced phase-space%
\begin{eqnarray}
\mathcal{\bar{H}}_{0}^{\left( VI\right) } &\approx &\tfrac{\left( k+1\right)
\left( k+1\right) !}{2\left( \tilde{a}-\left( -\right) ^{k}\right) }\left(
-\right) ^{k}\left( \pi _{i_{1}\cdots i_{k}\parallel j}^{2}-\pi
_{i_{1}\cdots i_{k}\parallel 0}^{2}\right)  \notag \\
&&+\tfrac{\left( -\right) ^{k}}{2\left( k+1\right) ^{2}\left( k+1\right) !}%
F^{i_{1}\cdots i_{k+1}\parallel j}\left( F_{i_{1}\cdots i_{k+1}\parallel j}-%
\tilde{a}F_{j\left[ i_{1}\cdots i_{k}\parallel i_{k+1}\right] }\right) .
\label{vic10a}
\end{eqnarray}%
From the expression in the above we conclude that also in this case
ghost-modes are involved.

Employing the same procedure as in the previous subsections, we derive for
the functional
\begin{equation}
S_{0}^{\left( VI\right) }\left[ A_{\mu _{1}\cdots \mu _{k}\parallel \alpha }%
\right] =\int \mathrm{d}^{D}x\mathcal{L}_{0}^{\left( VI\right) }
\label{vic11}
\end{equation}%
the generating set of gauge transformations%
\begin{equation}
\delta _{\epsilon ,\bar{\epsilon},\xi }^{\left( VI\right) }A_{\mu _{1}\cdots
\mu _{k}\parallel \alpha }=\partial _{\lbrack \mu _{1}}\epsilon _{\mu
_{2}\cdots \mu _{k}]\alpha }+\partial _{\lbrack \mu _{1}}\xi _{\mu
_{2}\cdots \mu _{k}]\mid \alpha }+\left( -\right) ^{k}\partial _{\alpha }%
\bar{\epsilon}_{\mu _{1}\cdots \mu _{k}},  \label{vic12}
\end{equation}%
where the bosonic gauge parameters of $\epsilon $-type are completely
antisymmetric.

Is is remarkable that also in this situation, the generating set of gauge
transformations (\ref{vic12}) is richer than the original one (\ref{3m1}).

\subsection{Case VII\label{VII}}

Here, the real parameters $a_{1}$ and $a_{2}$ possess the domains of values
precized in (\ref{c7vii}) and the corresponding Lagrangian density
\begin{equation}
\mathcal{L}_{0}^{\left( VII\right) }=\tfrac{1}{2\left( k+1\right) \left(
k+1\right) !}\left[ -\tfrac{\left( -\right) ^{k}}{k+1}\left( F_{\mu
_{1}\cdots \mu _{k+1}\parallel \alpha }\right) ^{2}+a_{1}F_{\mu _{1}\cdots
\mu _{k}\beta \parallel \alpha }F^{\mu _{1}\cdots \mu _{k}\alpha \parallel
\beta }+a_{2}\left( F_{\mu _{1}\cdots \mu _{k}}\right) ^{2}\right]
\label{viic1}
\end{equation}%
coincides with the original one (\ref{3m2}). In this case, the definitions (%
\ref{c1}) of the canonical momenta display the Abelian primary constraints (%
\ref{c2}) and lead [via the Lagrangian's Legendre transformation in respect
with the generalized velocities] to the canonical Hamiltonian density%
\begin{eqnarray}
\mathcal{H}_{0}^{\left( VII\right) } &=&-kA^{0i_{1}\cdots i_{k-1}\parallel
\mu }\left( \partial ^{l}\pi _{li_{1}\cdots i_{k-1}\parallel \mu }\right) +%
\tfrac{\left( -\right) ^{k}}{2\left( k+1\right) ^{2}\left( k+1\right) !}%
\left( F_{i_{1}\cdots i_{k+1}\parallel j}\right) ^{2}  \notag \\
&&+\tfrac{\left( k+1\right) \left( k+1\right) !}{2\left( a_{1}-\left(
-\right) ^{k}\right) }\pi ^{i_{1}\cdots i_{k}\parallel j}\left( \pi
_{i_{1}\cdots i_{k}\parallel j}-\tfrac{a_{1}}{ka_{1}+\left( -\right) ^{k}}%
\pi _{\left[ i_{1}\cdots i_{k}\parallel j\right] }\right.  \notag \\
&&\left. -\tfrac{a_{2}}{a_{1}+a_{2}\left( D-k\right) -\left( -\right) ^{k}}%
\pi _{\lbrack i_{1}\cdots i_{k-1}}^{\prime }\sigma _{i_{k]j}}^{\left.
{}\right. }\right)  \notag \\
&&+\tfrac{\left( k+1\right) \left( k+1\right) !}{2\left( a_{1}+a_{2}-\left(
-\right) ^{k}\right) }\left( \pi _{i_{1}\cdots i_{k}\parallel 0}\right) ^{2}+%
\tfrac{\left( -\right) ^{k}a_{1}}{ka_{1}+\left( -\right) ^{k}}\pi
_{i_{1}\cdots i_{k}\parallel j}F^{i_{1}\cdots i_{k}j\parallel 0}  \notag \\
&&-\tfrac{\left( -\right) ^{k}a_{2}}{a_{1}+a_{2}-\left( -\right) ^{k}}%
F_{i_{1}\cdots i_{k}}^{\prime }\left( \pi ^{i_{1}\cdots i_{k}\parallel 0}+%
\tfrac{\left( -\right) ^{k}a_{1}-1}{2\left( k+1\right) \left( k+1\right) !}%
F^{\prime i_{1}\cdots i_{k}}\right)  \notag \\
&&+\tfrac{1+\left( -\right) ^{k}ka_{1}-\left( k+1\right) a_{1}^{2}}{2\left(
k+1\right) ^{2}\left( k+1\right) !\left( ka_{1}+\left( -\right) ^{k}\right) }%
\left( F_{i_{1}\cdots i_{k+1}\parallel 0}\right) ^{2}-\tfrac{a_{1}}{2\left(
k+1\right) \left( k+1\right) !}F^{i_{1}\cdots i_{k}j\parallel
l}F_{i_{1}\cdots i_{k}l\parallel j}.  \label{viic2}
\end{eqnarray}

At this stage, if we use the first-class character of the primary
constraints (\ref{c2}), then, by asking them the time preservation, we get%
\begin{equation}
\left[ G_{i_{1}\cdots i_{k-1}}^{(1)},H_{0}^{\left( VII\right) }\right]
=G_{i_{1}\cdots i_{k-1}}^{(2)}\approx 0,\quad \left[ G_{i_{1}\cdots
i_{k-1}\parallel j}^{(1)},H_{0}^{\left( VII\right) }\right] =G_{i_{1}\cdots
i_{k-1}\parallel j}^{(2)}\approx 0.  \label{viic3}
\end{equation}%
results that display the secondary constraints $G_{i_{1}\cdots
i_{k-1}}^{(2)} $ and $G_{i_{1}\cdots i_{k-1}\parallel j}^{(2)}$ [whose
concrete expressions are respectively written in (\ref{iiic5a}) and (\ref%
{iiic5b})] .The dependence of the functions (\ref{c2}), (\ref{iiic5a}) and (%
\ref{iiic5b}) only on the canonical momenta lead to their Abelian character
in the Poisson brackets. In this light, the consistency of the secondary
constraints reduces to the computations of the Poisson brackets between them
and the canonical Haniltonian. It can be checked that
\begin{equation}
\left[ G_{i_{1}\cdots i_{k-1}}^{(2)},H_{0}^{\left( VII\right) }\right] =0=%
\left[ G_{i_{1}\cdots i_{k-1}\parallel j}^{(2)},H_{0}^{\left( VII\right) }%
\right]  \label{viic4}
\end{equation}%
so the Dirac algorithm stops at this level.

Putting the results (\ref{viic3}) and (\ref{viic4}), we conclude that the
canonical Hamiltonian (\ref{viic2}) is just the first-class Hamiltonian of
the system. In view of counting the number of independent degrees of
freedom, we inovke the irreducible character of the first-class constraints (%
\ref{c2}) supplemented with the $L=k-1$ reducibility functions (\ref{iic7a}%
)--(\ref{iic7b}) of the secondary first-class constraints $G_{i_{1}\cdots
i_{k-1}\parallel j}^{(2)}$ and $G_{i_{1}\cdots i_{k-1}\parallel j}^{(2)}$
and get%
\begin{equation}
N_{DOF}^{\left( VII\right) }=D\left(
\begin{array}{c}
D-2 \\
k%
\end{array}%
\right) .  \label{viic5}
\end{equation}

As in the other six situations, we are interested whether all the
independent degrees of freedom are physical. In view of this, we take the
canonical gauge conditions (\ref{ic10a}), (\ref{iic9d}) and (\ref{iic9e})
and then we evaluate the restriction of the firs-class Hamiltonian on the
cooresponding reduced phase-space. Simple calculations reveal%
\begin{eqnarray}
\mathcal{\bar{H}}_{0}^{\left( VII\right) } &\approx &\tfrac{\left(
k+1\right) \left( k+1\right) !}{2\left( 1-\left( -\right) ^{k}a_{1}\right) }%
\left( \pi _{i_{1}\cdots i_{k}\parallel j}^{2}-\tfrac{ka_{2}}{%
a_{1}+a_{2}\left( D-k\right) -\left( -\right) ^{k}}\pi _{i_{1}\cdots
i_{k-1}}^{\prime 2}\right.  \notag \\
&&\left. -\tfrac{a_{1}}{\left( k+1\right) \left( ka_{1}+\left( -\right)
^{k}\right) }p_{i_{1}\cdots i_{k+1}}^{2}-\tfrac{a_{1}-\left( -\right) ^{k}}{%
a_{1}+a_{2}-\left( -\right) ^{k}}p_{i_{1}\cdots i_{k}}^{2}\right)  \notag \\
&&-\tfrac{a_{2}}{2\left( k+1\right) \left( k+1\right) !}\left(
F_{i_{1}\cdots i_{k}}^{\prime }\right) ^{2}+\tfrac{1+\left( -\right)
^{k}\left( k-1\right) a_{1}-ka_{1}^{2}}{2\left( k+1\right) ^{2}\left(
k+1\right) !\left( ka_{1}+\left( -\right) ^{k}\right) }\left( F_{i_{1}\cdots
i_{k+1}\parallel 0}\right) ^{2}  \notag \\
&&+\tfrac{\left( -\right) ^{k}}{2\left( k+1\right) ^{2}\left( k+1\right) !}%
F^{i_{1}\cdots i_{k+1}\parallel j}\left( F_{i_{1}\cdots i_{k+1}\parallel
j}-a_{1}F_{j\left[ i_{1}\cdots i_{k}\parallel i_{k+1}\right] }\right) ,
\label{viic5a}
\end{eqnarray}%
where we employed the notations%
\begin{eqnarray}
p_{i_{1}\cdots i_{k+1}} &\equiv &\pi _{\left[ i_{1}\cdots i_{k}\parallel
i_{k+1}\right] }+\tfrac{1-\left( -\right) ^{k}a_{1}}{\left( k+1\right)
\left( k+1\right) !}F_{i_{1}\cdots i_{k+1}\parallel 0},  \label{viic5b} \\
p_{i_{1}\cdots i_{k}} &\equiv &\pi _{i_{1}\cdots i_{k}\parallel 0}-\tfrac{%
\left( -\right) ^{k}a_{2}}{\left( k+1\right) \left( k+1\right) !}%
F_{i_{1}\cdots i_{k}}^{\prime }.  \label{viic5c}
\end{eqnarray}%
At this stage we can state that ghost modes are absent from (\ref{viic5a})
iff the real constants $a_{1}$ and $a_{2}$ are subjects of the inqualities%
\begin{equation}
\tfrac{a_{2}}{a_{1}+a_{2}\left( D-k\right) -\left( -\right) ^{k}}<0,\quad
\tfrac{a_{1}}{ka_{1}+\left( -\right) ^{k}}<0,\quad \tfrac{a_{1}-\left(
-\right) ^{k}}{a_{1}+a_{2}-\left( -\right) ^{k}}<0.  \label{viic5d}
\end{equation}

By analyzing the inequalities (\ref{viic5d}) we establish their
incompatibility, result that imeadiately imply the ghost modes are also
present in this situation.

Finally, if we return to the Lagrangian formulation [via extended action],
we derive for the functional
\begin{equation}
S_{0}^{\left( VII\right) }\left[ A_{\mu _{1}\cdots \mu _{k}\parallel \alpha }%
\right] =\int \mathrm{d}^{D}x\mathcal{L}_{0}^{\left( VII\right) }
\label{viic6}
\end{equation}%
the initial generating set of gauge transformations (\ref{3m1}).

\section{First-order formulations\label{FO}}

In the present part we will derive the first-order formulations associated
with the second-order models previously investigated. These constructions
can be done using some auxiliary matter/gauge fields. More precisely, we
shall show that: i) if the second-order Lagrangian can be written as a
bilinear combination of some classical observables that are linearly in the
field-strengths $F_{\mu _{1}\cdots \mu _{k+1}\parallel \alpha }$ then the
corresponding first-order formulation requires only auxiliary matter fields;
ii) if the second-order Lagrangian cannot be written in terms of the
classical observables that are linearly in the field-strengths $F_{\mu
_{1}\cdots \mu _{k+1}\parallel \alpha }$ then the associated first-order
Lagrangian involves some auxiliary gauge fields.

The program for constructing the first-order Lagrangian $\overline{\mathcal{L%
}}_{0}$ corresponding to the generic second-order one%
\begin{equation}
\mathcal{L}_{0}=\mathcal{L}_{0}\left( \left[ A_{\mu _{1}\cdots \mu
_{k}\parallel \alpha }\right] \right) =\frac{1}{2}F_{\mu _{1}\cdots \mu
_{k+1}\parallel \alpha }M_{\nu _{1}\cdots \nu _{k+1}\parallel \beta }^{\mu
_{1}\cdots \mu _{k+1}\parallel \alpha }F^{\nu _{1}\cdots \nu _{k+1}\parallel
\beta },  \label{fo1}
\end{equation}%
where $M_{\nu _{1}\cdots \nu _{k+1}\parallel \beta }^{\mu _{1}\cdots \mu
_{k+1}\parallel \alpha }$ is a Lorentz nonderivative constant tensor,
consists in the following steps:

\begin{enumerate}
\item[i)] one postulates for $\overline{\mathcal{L}}_{0}$ the gauge
transformations
\begin{equation}
\bar{\delta}_{\epsilon ,\xi }A_{\mu _{1}\cdots \mu _{k}\parallel \alpha
}\equiv \delta _{\epsilon ,\xi }A_{\mu _{1}\cdots \mu _{k}\parallel \alpha },
\label{fo2}
\end{equation}%
where $\delta A_{\mu _{1}\cdots \mu _{k}\parallel \alpha }$ is the
generating set of gauge transformations corresponding to $\mathcal{L}_{0}$.

\item[ii)] one computes the gauge variations of the field-strengths $F_{\mu
_{1}\cdots \mu _{k+1}\parallel \alpha }$ under (\ref{fo2}) and looks for
linear, nonderivative combinations $\mathcal{F}_{\mu _{1}\cdots \mu
_{k+1}\alpha }$ of $F_{\mu _{1}\cdots \mu _{k+1}\parallel \alpha }$ that are
invariant under (\ref{fo2}).

\item[iii)] if there are the tensors $\mathcal{F}_{\mu _{1}\cdots \mu
_{k+1}\alpha }$ such that the Lagrangian density (\ref{fo1}) can be written
only in terms of them, then the first-order Lagrangian $\overline{\mathcal{L}%
}_{0}$ is obtained with the help of some bosonic matter fields $\mathcal{B}%
_{\mu _{1}\cdots \mu _{k+1}\alpha }$ [that display the symmetry properties
of $\mathcal{F}_{\mu _{1}\cdots \mu _{k+1}\alpha }$].

\item[iv)] if there are no such tensors, then the first-order Lagrangian $%
\overline{\mathcal{L}}_{0}$ can be written as a quadratic form in an
auxiliary gauge field $\omega _{\alpha \parallel \mu _{1}\cdots \mu _{k+1}}$
[$\omega _{\alpha \parallel \mu _{1}\cdots \mu _{k+1}}=\frac{1}{\left(
k+1\right) !}\omega _{\alpha \parallel \left[ \mu _{1}\cdots \mu _{k+1}%
\right] }$] with the gauge transformations specified.
\end{enumerate}

In the sequel we shall apply this program for each of the seven situations
analyzed in the previous subsection.

\subsection{Case I \label{foI}}

In this situation, the Lagrangian density (\ref{ic1}) can be written in
terms of the gauge-invariant objects $F_{\left[ \mu _{1}\cdots \mu
_{k+1}\parallel \alpha \right] }$ as%
\begin{equation}
\mathcal{L}_{0}^{\left( I\right) }\equiv \tfrac{\left( -\right) ^{k+1}}{%
2\left( k+1\right) ^{2}\left( k+2\right) !}\left( F_{\left[ \mu _{1}\cdots
\mu _{k+1}\parallel \alpha \right] }\right) ^{2}.  \label{ifo1}
\end{equation}%
Therefore, the corresponding first-order Lagrangian, $\mathcal{\bar{L}}%
_{0}^{\left( I\right) }$, depends on the original gauge field [through the
combination $F_{\left[ \mu _{1}\cdots \mu _{k+1}\parallel \alpha \right] }$]
and the matter $\left( k+2\right) $-form $B_{\mu _{1}\cdots \mu _{k+2}}$%
\begin{eqnarray}
\overline{\mathcal{L}}_{0}^{\left( I\right) } &\equiv &\overline{\mathcal{L}}%
_{0}^{\left( I\right) }\left( B_{\mu _{1}\cdots \mu _{k+2}},F_{\mu
_{1}\cdots \mu _{k+1}\parallel \alpha }\right)  \notag \\
&=&B_{\mu _{1}\cdots \mu _{k+2}}\left( \tfrac{1}{k+1}F^{\left[ \mu
_{1}\cdots \mu _{k+1}\parallel \mu _{k+2}\right] }-\left( -\right) ^{k}%
\tfrac{\left( k+2\right) !}{2}B^{\mu _{1}\cdots \mu _{k+2}}\right) .
\label{ifo2}
\end{eqnarray}%
A generating set of gauge transformations for the first-order action%
\begin{equation}
\bar{S}_{0}^{\left( I\right) }\left[ A_{\mu _{1}\cdots \mu _{k}\parallel
\alpha },B_{\mu _{1}\cdots \mu _{k+2}}\right] =\int \mathrm{d}^{D}x\overline{%
\mathcal{L}}_{0}^{\left( I\right) }  \label{ifo3}
\end{equation}%
consists in
\begin{equation}
\bar{\delta}_{\epsilon ,\xi }^{\left( I\right) }A_{\mu _{1}\cdots \mu
_{k}\parallel \alpha }=\delta _{\epsilon ,\xi }^{\left( I\right) }A_{\mu
_{1}\cdots \mu _{k}\parallel \alpha },\quad \bar{\delta}_{\epsilon ,\xi
}^{\left( I\right) }B_{\mu _{1}\cdots \mu _{k+2}}=0,  \label{ifo4}
\end{equation}%
where the gauge transformations $\delta _{\epsilon ,\xi }^{\left( I\right)
}A_{\mu _{1}\cdots \mu _{k}\parallel \alpha }$ have the concrete form (\ref%
{ic11}).

It is easy to see that the elimination of the auxiliary variables $B_{\mu
_{1}\cdots \mu _{k+2}}$ on their own field equations%
\begin{equation}
\frac{\delta \overline{\mathcal{L}}_{0}^{\left( I\right) }}{\delta B_{\mu
_{1}\cdots \mu _{k+2}}}\equiv \tfrac{1}{k+1}F^{\left[ \mu _{1}\cdots \mu
_{k+1}\parallel \mu _{k+2}\right] }-\left( -\right) ^{k}\left( k+2\right)
!B^{\mu _{1}\cdots \mu _{k+2}}=0  \label{ifo5x}
\end{equation}%
leads to%
\begin{equation}
B_{\mu _{1}\cdots \mu _{k+2}}=\tfrac{\left( -\right) ^{k}}{\left( k+1\right)
\left( k+2\right) !}F_{\left[ \mu _{1}\cdots \mu _{k+1}\parallel \mu _{k+2}%
\right] }.  \label{ifo5}
\end{equation}%
Inserting the solution (\ref{ifo5}) into (\ref{ifo2}), we regain the
second-order formulation%
\begin{equation}
\mathcal{L}_{0}^{\left( I\right) }=\left. \overline{\mathcal{L}}_{0}^{\left(
I\right) }\right\vert _{B_{\mu _{1}\cdots \mu _{k+2}}=\tfrac{\left( -\right)
^{k}}{\left( k+1\right) \left( k+2\right) !}F_{\left[ \mu _{1}\cdots \mu
_{k+1}\parallel \mu _{k+2}\right] }}.  \label{ifo6}
\end{equation}

\subsection{Case II \label{foII}}

Following the program exposed in the beginning of this section, we firstly
compute the gauge variation of the field-strength $F_{\mu _{1}\cdots \mu
_{k+1}\parallel \alpha }$ under the gauge transformations (\ref{fo2})
corresponding to the analyzed situation (\ref{iic11})%
\begin{equation}
\bar{\delta}_{\epsilon ,\varepsilon }^{\left( II\right) }F_{\mu _{1}\cdots
\mu _{k+1}\parallel \alpha }=\left( -\right) ^{k+1}k\partial _{\alpha
}\left( \partial _{\lbrack \mu _{1}}\epsilon _{\mu _{2}\cdots \mu
_{k+1}]}\right) +\partial _{\lbrack \mu _{1}}\epsilon _{\mu _{2}\cdots \mu
_{k+1}]\alpha }.  \label{iifo1}
\end{equation}%
From the results (\ref{iifo1}) one can see that there are no combinations of
$\mathcal{F}_{\mu _{1}\cdots \mu _{k+1}\alpha }$-type. In view of these, the
first-order formulation associated to this limit case, also addressed in
\cite{zinoviev} for $k=1$ and $k=2$, can be done with the help of an
auxiliary gauge field $\omega _{\alpha \parallel \mu _{1}\cdots \mu _{k+1}}$
[$\omega _{\alpha \parallel \mu _{1}\cdots \mu _{k+1}}=\frac{1}{\left(
k+1\right) !}\omega _{\alpha \parallel \left[ \mu _{1}\cdots \mu _{k+1}%
\right] }$] introduced in order to compensate the gauge variation of the
field-strength $F_{\mu _{1}\cdots \mu _{k+1}\parallel \alpha }$ (\ref{iifo1}%
). We postulate for the aforementioned gauge field the gauge transformations
\begin{equation}
\bar{\delta}_{\epsilon ,\xi }^{\left( II\right) }\omega _{\alpha \parallel
\mu _{1}\cdots \mu _{k+1}}=\partial _{\alpha }\left( \epsilon _{\mu
_{1}\cdots \mu _{k+1}}-k\partial _{\lbrack \mu _{1}}\epsilon _{\mu
_{2}\cdots \mu _{k+1}]}\right)  \label{iifo2}
\end{equation}%
that mimic (\ref{iifo1}). At this level, one can construct the
gauge-invariant tensors%
\begin{equation}
I_{\mu _{1}\cdots \mu _{k+1}\parallel \alpha }\equiv F_{\mu _{1}\cdots \mu
_{k+1}\parallel \alpha }-\omega _{\left[ \mu _{1}\parallel \mu _{2}\cdots
\mu _{k+1}\right] \alpha },  \label{iifo3}
\end{equation}%
antisymmetric in the first two Lorentz indices, useful in order to write
down the first-order Lagrangian density.

The first-order Lagrangian $\overline{\mathcal{L}}_{0}^{\left( II\right) }$
is obtained subtracting from the second-order Lagrangian $\mathcal{L}%
_{0}^{\left( II\right) }$ the quadratic combinations in the gauge-invariant
tensors (\ref{iifo3}) that cancel the terms of $\mathcal{L}_{0}^{\left(
II\right) }$%
\begin{eqnarray}
\overline{\mathcal{L}}_{0}^{\left( II\right) } &=&\mathcal{L}_{0}^{\left(
II\right) }+\tfrac{\left( -\right) ^{k+1}}{2\left( k+1\right) \left(
k+1\right) !}I_{\mu _{1}\cdots \mu _{k+1}\parallel \alpha }\left( \tfrac{1}{%
k+1}I^{\mu _{1}\cdots \mu _{k+1}\parallel \alpha }\right.  \notag \\
&&\left. +\tfrac{1}{k}I^{\mu _{1}\cdots \mu _{k}\alpha \parallel \mu _{k+1}}-%
\tfrac{1}{k}I^{[\mu _{1}\cdots \mu _{k}}\sigma ^{\mu _{k+1}]\alpha }\right)
\label{iifo4}
\end{eqnarray}%
where we used the notations
\begin{equation}
I_{\mu _{1}\cdots \mu _{k}}\equiv \sigma ^{\mu _{k+1}\alpha }I_{\mu
_{1}\cdots \mu _{k+1}\parallel \alpha }.  \label{iifo5}
\end{equation}

The manner just has employed for constructing the first-order Lagrangian (%
\ref{iifo4}) allow us to conclude that:

\begin{enumerate}
\item[a)] The local density (\ref{iifo4}) is gauge invariant under the gauge
transformations (\ref{fo2}) [with the right-hand side replaced by the
expression (\ref{iic11})] and (\ref{iifo2});

\item[b)] Eliminating in (\ref{iifo4}) the auxiliary variable on their own
field equations one obtains the second-order Lagrangian (\ref{iic1}).
\end{enumerate}

Whether the first conclusion is obvious, the second one will become
transparent as follows. Inserting the definitions (\ref{ifo3}) into the
formula (\ref{iifo4}) one derives the concrete expression of the first-order
Lagrangian density%
\begin{eqnarray}
\overline{\mathcal{L}}_{0}^{\left( II\right) } &\equiv &\overline{\mathcal{L}%
}_{0}^{\left( II\right) }\left( \omega _{\alpha \parallel \mu _{1}\cdots \mu
_{k+1}},F_{\mu _{1}\cdots \mu _{k+1}\parallel \alpha }\right)  \notag \\
&=&\tfrac{1}{k\left( k+1\right) \left( k+1\right) !}\omega ^{\alpha
\parallel \mu _{1}\cdots \mu _{k+1}}\left[ \tfrac{1}{2}\left( \omega _{\left[
\mu _{1}\parallel \mu _{2}\cdots \mu _{k+1}\right] \alpha }-\omega _{\lbrack
\mu _{1}\cdots \mu _{k}}\sigma _{\mu _{k+1}]\alpha }\right) \right.  \notag
\\
&&\left. -F_{\mu _{1}\cdots \mu _{k+1}\parallel \alpha }+F_{[\mu _{1}\cdots
\mu _{k}}\sigma _{\mu _{k+1}]\alpha }\right] ,  \label{iifo6}
\end{eqnarray}%
where we employed the notations%
\begin{equation}
\omega _{\mu _{1}\cdots \mu _{k}}\equiv \sigma ^{\alpha \beta }\omega
_{\alpha \parallel \beta \mu _{1}\cdots \mu _{k}}  \label{iifo7}
\end{equation}%
By direct computation, one infers the field equations%
\begin{eqnarray*}
\frac{\delta \overline{\mathcal{L}}_{0}^{\left( II\right) }}{\delta \omega
^{\alpha \parallel \mu _{1}\cdots \mu _{k+1}}} &\equiv &\tfrac{1}{k\left(
k+1\right) \left( k+1\right) !}\left[ \omega _{\left[ \mu _{1}\parallel \mu
_{2}\cdots \mu _{k+1}\right] \alpha }-\omega _{\lbrack \mu _{1}\cdots \mu
_{k}}\sigma _{\mu _{k+1}]\alpha }\right. \\
\left. -F_{\mu _{1}\cdots \mu _{k+1}\parallel \alpha }+F_{[\mu _{1}\cdots
\mu _{k}}\sigma _{\mu _{k+1}]\alpha }\right] &=&0
\end{eqnarray*}%
whose solutions read as%
\begin{equation}
\omega _{\alpha \parallel \mu _{1}\cdots \mu _{k+1}}=\left( -\right)
^{k}\left( F_{\mu _{1}\cdots \mu _{k+1}\parallel \alpha }-\tfrac{1}{k+1}F_{%
\left[ \mu _{1}\cdots \mu _{k+1}\parallel \alpha \right] }\right) .
\label{iifo9}
\end{equation}%
Inserting the results (\ref{iifo9}) into the first-order Lagrangian density (%
\ref{iifo6}) we establish%
\begin{equation}
\mathcal{L}_{0}^{\left( II\right) }=\left. \overline{\mathcal{L}}%
_{0}^{\left( II\right) }\right\vert _{\omega _{\alpha \parallel \mu
_{1}\cdots \mu _{k+1}}=\left( -\right) ^{k}\left( F_{\mu _{1}\cdots \mu
_{k+1}\parallel \alpha }-\tfrac{1}{k+1}F_{\left[ \mu _{1}\cdots \mu
_{k+1}\parallel \alpha \right] }\right) }.  \label{iifo10}
\end{equation}

\subsection{Case III \label{foIII}}

This case is similar to the previous one in the sense that the derivation of
the first-order Lagrangian density requires some auxiliary gauge fields.
This is due to the absence of the gauge-invariant $\mathcal{F}_{\mu
_{1}\cdots \mu _{k+1}\alpha }$-type tensors. Precisely, if we compute the
gauge variation of the field-strength $F_{\mu _{1}\cdots \mu _{k+1}\parallel
\alpha }$ under the gauge transformations (\ref{fo2}) corresponding to the
analyzed situation (\ref{iiic12})%
\begin{equation}
\bar{\delta}_{\epsilon ,\varepsilon }^{\left( III\right) }F_{\mu _{1}\cdots
\mu _{k+1}\parallel \alpha }=\partial _{\lbrack \mu _{1}}\epsilon _{\mu
_{2}\cdots \mu _{k}}\sigma _{\mu _{k+1}]\alpha }+\partial ^{\sigma }\partial
_{\lbrack \mu _{1}}\epsilon _{\mu _{2}\cdots \mu _{k+1}]\alpha \sigma },
\label{iiifo1}
\end{equation}%
we observe that we cannot identify any linearly and nonderivative
combination of $F_{\mu _{1}\cdots \mu _{k+1}\parallel \alpha }$ that is
gauge-invariant. Nevertheless, there are some linearly and nonderivative
combinations of $F_{\mu _{1}\cdots \mu _{k+1}\parallel \alpha }$
\begin{equation}
\bar{F}_{\mu _{1}\cdots \mu _{k+1}\parallel \alpha }\equiv F_{\mu _{1}\cdots
\mu _{k+1}\parallel \alpha }-\tfrac{1}{D-k}F_{[\mu _{1}\cdots \mu
_{k}}\sigma _{\mu _{k+1}]\alpha },  \label{iiifo2}
\end{equation}%
whose gauge variations do not depend on the $\left( k-1\right) $-form gauge
parameter%
\begin{equation}
\bar{\delta}_{\epsilon ,\varepsilon }^{\left( III\right) }\bar{F}_{\mu
_{1}\cdots \mu _{k+1}\parallel \alpha }=\partial ^{\sigma }\partial
_{\lbrack \mu _{1}}\epsilon _{\mu _{2}\cdots \mu _{k+1}]\alpha \sigma }.
\label{iiifo3}
\end{equation}%
The definitions (\ref{iiifo2}) reveal the traceless character of the
antisymmetric tensors $\bar{F}_{\mu _{1}\cdots \mu _{k+1}\parallel \alpha }$
[$\bar{F}_{\mu _{1}\cdots \mu _{k+1}\parallel \alpha }=\frac{1}{\left(
k+1\right) !}\bar{F}_{\left[ \mu _{1}\cdots \mu _{k+1}\right] \parallel
\alpha }$]%
\begin{equation}
\sigma ^{\mu _{k+1}\alpha }\bar{F}_{\mu _{1}\cdots \mu _{k+1}\parallel
\alpha }=0.  \label{iiifo4}
\end{equation}

In order to construct the first-order Lagrangian density, we introduce the
auxiliary gauge fields $\bar{\omega}_{\alpha \parallel \mu _{1}\cdots \mu
_{k+1}}$ with the algebraic properties of the tensors $\bar{F}_{\mu
_{1}\cdots \mu _{k+1}\parallel \alpha }$ [$\bar{\omega}_{\alpha \parallel
\mu _{1}\cdots \mu _{k+1}}=\frac{1}{\left( k+1\right) !}\bar{\omega}_{\alpha
\parallel \left[ \mu _{1}\cdots \mu _{k+1}\right] }$, $\sigma ^{\alpha \mu
_{1}}\bar{\omega}_{\alpha \parallel \mu _{1}\cdots \mu _{k+1}}=0$] which are
subject to the gauge transformations%
\begin{equation}
\bar{\delta}_{\epsilon ,\varepsilon }^{\left( III\right) }\bar{\omega}%
_{\alpha \parallel \mu _{1}\cdots \mu _{k+1}}=\partial _{\alpha }\partial
^{\sigma }\epsilon _{\mu _{1}\cdots \mu _{k+1}\sigma }.  \label{iiifo5}
\end{equation}%
Based on the results (\ref{iiifo3}), the newly introduced tensor gauge
fields $\bar{\omega}_{\alpha \parallel \mu _{1}\cdots \mu _{k+1}}$ allows to
identify the gauge-invariant Lorentz tensors%
\begin{equation}
\bar{I}_{\mu _{1}\cdots \mu _{k+1}\parallel \alpha }\equiv \bar{F}_{\mu
_{1}\cdots \mu _{k+1}\parallel \alpha }-\bar{\omega}_{\left[ \mu
_{1}\parallel \mu _{2}\cdots \mu _{k+1}\right] \alpha }  \label{iiifo6}
\end{equation}%
that satisfy%
\begin{equation}
\bar{I}_{\mu _{1}\cdots \mu _{k+1}\parallel \alpha }=\tfrac{1}{\left(
k+1\right) !}\bar{I}_{\left[ \mu _{1}\cdots \mu _{k+1}\right] \parallel
\alpha },\quad \sigma ^{\mu _{k+1}\alpha }\bar{I}_{\mu _{1}\cdots \mu
_{k+1}\parallel \alpha }=0.  \label{iiifo7}
\end{equation}

Proceeding as in the previous case, the first-order Lagrangian density $%
\overline{\mathcal{L}}_{0}^{\left( III\right) }$ can be written as the sum
between the second-order Lagrangian $\mathcal{L}_{0}^{\left( III\right) }$
and some quadratic combinations in the gauge-invariant tensors (\ref{iiifo6}%
) that cancel the terms of $\mathcal{L}_{0}^{\left( III\right) }$%
\begin{equation}
\overline{\mathcal{L}}_{0}^{\left( III\right) }=\mathcal{L}_{0}^{\left(
III\right) }+\tfrac{\left( -\right) ^{k}}{2\left( k+1\right) \left(
k+1\right) !}\bar{I}_{\mu _{1}\cdots \mu _{k+1}\parallel \alpha }\left(
\tfrac{1}{k+1}\bar{I}^{\mu _{1}\cdots \mu _{k+1}\parallel \alpha }+\tfrac{1}{%
k}\bar{I}^{\mu _{1}\cdots \mu _{k}\alpha \parallel \mu _{k+1}}\right)
\label{iiifo8}
\end{equation}%
By construction, the Lagrangian density is manifestly gauge-invariaant under
the gauge transformations (\ref{fo2}) [with the right-hand side replaced by
the expression (\ref{iiic12})] and (\ref{iiifo5}). Also, this reduces to the
second-order Lagrangian (\ref{iiic1}) by eliminating the auxiliary fields $%
\bar{\omega}_{\alpha \parallel \mu _{1}\cdots \mu _{k+1}}$ on their own
field equations as we shall see.

Inserting the definitions (\ref{iiifo6}) into the relation (\ref{iiifo8})
one obtains the concrete expression of the first-order Lagrangian density%
\begin{eqnarray}
\overline{\mathcal{L}}_{0}^{\left( III\right) } &\equiv &\overline{\mathcal{L%
}}_{0}^{\left( III\right) }\left( \bar{\omega}_{\alpha \parallel \mu
_{1}\cdots \mu _{k+1}},F_{\mu _{1}\cdots \mu _{k+1}\parallel \alpha }\right)
\notag \\
&=&\tfrac{1}{\left( k+1\right) \left( k+1\right) !}\bar{\omega}^{\alpha
\parallel \mu _{1}\cdots \mu _{k+1}}\left( \tfrac{1}{2}\bar{\omega}_{\left[
\mu _{1}\parallel \mu _{2}\cdots \mu _{k+1}\right] \alpha }-\bar{F}_{\mu
_{1}\cdots \mu _{k+1}\parallel \alpha }\right) .  \label{iiifo9}
\end{eqnarray}%
The field equations in respect with the auxiliary variables are%
\begin{equation}
\frac{\delta \overline{\mathcal{L}}_{0}^{\left( III\right) }}{\delta \bar{%
\omega}^{\alpha \parallel \mu _{1}\cdots \mu _{k+1}}}\equiv \tfrac{1}{%
k\left( k+1\right) \left( k+1\right) !}\left( \bar{F}_{\mu _{1}\cdots \mu
_{k+1}\parallel \alpha }-\bar{\omega}_{\left[ \mu _{1}\parallel \mu
_{2}\cdots \mu _{k+1}\right] \alpha }\right) =0  \label{iiifo10}
\end{equation}%
whose solutions read as%
\begin{equation}
\bar{\omega}_{\alpha \parallel \mu _{1}\cdots \mu _{k+1}}=\left( -\right)
^{k}\left( \bar{F}_{\mu _{1}\cdots \mu _{k+1}\parallel \alpha }-\tfrac{1}{k+1%
}\bar{F}_{\left[ \mu _{1}\cdots \mu _{k+1}\parallel \alpha \right] }\right) .
\label{iiifo11}
\end{equation}%
Inserting the results (\ref{iiifo10}) into the first-order Lagrangian
density (\ref{iiifo9}) we establish%
\begin{equation}
\mathcal{L}_{0}^{\left( III\right) }=\left. \overline{\mathcal{L}}%
_{0}^{\left( III\right) }\right\vert _{\bar{\omega}_{\alpha \parallel \mu
_{1}\cdots \mu _{k+1}}=\left( -\right) ^{k}\left( \bar{F}_{\mu _{1}\cdots
\mu _{k+1}\parallel \alpha }-\tfrac{1}{k+1}\bar{F}_{\left[ \mu _{1}\cdots
\mu _{k+1}\parallel \alpha \right] }\right) }.  \label{iiifo12}
\end{equation}

\subsection{Case IV \label{foIV}}

The present case is, in some sense, a mixing situation of the cases analyzed
in \ref{foI} and \ref{foIII}. Precisely, there are some gauge-invariant
tensors of $\mathcal{F}_{\mu _{1}\cdots \mu _{k+1}\alpha }$-type but,
however, the second-order Lagrangian density (\ref{ivc1}) cannot be written
in terms of them as a bilinear combinations (\ref{fo1}). Nevertheless, the
local function (\ref{ivc1}) can be represented as the sum between the
second-order Lagrangian density (\ref{iiic1}) and some non-trivial terms
that are quadratic in the aforementioned gauge-invariant tensors%
\begin{equation}
\mathcal{L}_{0}^{\left( IV\right) }=\mathcal{L}_{0}^{\left( III\right) }+%
\tfrac{1}{2\left( k+1\right) \left( k+1\right) !}\left( a-\left( -\right)
^{k}\tfrac{k+1}{k\left( D-k\right) }\right) F_{\mu _{1}\cdots \mu
_{k}}F^{\mu _{1}\cdots \mu _{k}}.  \label{ivfo1}
\end{equation}

Each of the terms in the right-hand side of the decomposition (\ref{ivfo1})
is gauge-invariant under the gauge transformations (\ref{fo2}) associated to
the case under discussion (\ref{ivc7}). Indeed, using the definitions (\ref%
{3m1b}) we derive the gauge variations of the field-strength $F_{\mu
_{1}\cdots \mu _{k+1}\parallel \alpha }$ under the gauge transformations (%
\ref{ivc7})%
\begin{equation}
\bar{\delta}_{\epsilon ,\varepsilon }^{\left( IV\right) }F_{\mu _{1}\cdots
\mu _{k+1}\parallel \alpha }=\partial ^{\sigma }\partial _{\lbrack \mu
_{1}}\epsilon _{\mu _{2}\cdots \mu _{k+1}]\alpha \sigma },  \label{ivfo3}
\end{equation}%
that lead to the gauge-invariance of the trace $F_{\mu _{1}\cdots \mu _{k}}$%
\begin{equation}
\bar{\delta}_{\epsilon ,\varepsilon }^{\left( IV\right) }F_{\mu _{1}\cdots
\mu _{k}}=0.  \label{ivfo4}
\end{equation}

The previous discussion supplemented with the results established in the
subsections \ref{foI} and \ref{foIII} allow us to linearize the second-order
Lagrangian density (\ref{ivfo1}) with the help of the auxiliary tensor
fields $\bar{\omega}_{\alpha \parallel \mu _{1}\cdots \mu _{k+1}}$%
\begin{equation}
\bar{\omega}_{\alpha \parallel \mu _{1}\cdots \mu _{k+1}}=\tfrac{1}{\left(
k+1\right) !}\bar{\omega}_{\alpha \parallel \left[ \mu _{1}\cdots \mu _{k+1}%
\right] },\quad \sigma ^{\alpha \mu _{1}}\bar{\omega}_{\alpha \parallel \mu
_{1}\cdots \mu _{k+1}}=0  \label{ivfo5}
\end{equation}%
and $B_{\mu _{1}\cdots \mu _{k}}$ [$B_{\mu _{1}\cdots \mu _{k}}=\frac{1}{k!}%
B_{\left[ \mu _{1}\cdots \mu _{k}\right] }$] subject to the gauge
transformations%
\begin{equation}
\bar{\delta}_{\epsilon ,\varepsilon }^{\left( IV\right) }\bar{\omega}%
_{\alpha \parallel \mu _{1}\cdots \mu _{k+1}}=\partial _{\alpha }\partial
^{\sigma }\epsilon _{\mu _{1}\cdots \mu _{k+1}\sigma },\quad \bar{\delta}%
_{\epsilon ,\varepsilon }^{\left( IV\right) }B_{\mu _{1}\cdots \mu _{k}}=0.
\label{ivfo6}
\end{equation}%
The auxiliary gauge fields $\bar{\omega}_{\alpha \parallel \mu _{1}\cdots
\mu _{k+1}}$ are responsible with the linearization of the first term in the
right-hand side of the decomposition (\ref{ivfo1}) while the auxiliary
matter vector field $B_{\mu _{1}\cdots \mu _{k}}$ is used to build the
first-order formulation associated with the second term in the right-hand
side decomposition (\ref{ivfo1}).

Invoking again the procedures developed in the subsections \ref{foI} and \ref%
{foIII}, we can write down the first-order Lagrangian density corresponding
to the second-order one (\ref{ivfo1})%
\begin{eqnarray}
\overline{\mathcal{L}}_{0}^{\left( IV\right) } &\equiv &\overline{\mathcal{L}%
}_{0}^{\left( IV\right) }\left( \bar{\omega}_{\alpha \parallel \mu
_{1}\cdots \mu _{k+1}},B_{\mu _{1}\cdots \mu _{k}},F_{\mu _{1}\cdots \mu
_{k+1}\parallel \alpha }\right)  \notag \\
&=&\tfrac{1}{\left( k+1\right) \left( k+1\right) !}\bar{\omega}^{\alpha
\parallel \mu _{1}\cdots \mu _{k+1}}\left( \tfrac{1}{2}\bar{\omega}_{\left[
\mu _{1}\parallel \mu _{2}\cdots \mu _{k+1}\right] \alpha }-\bar{F}_{\mu
_{1}\cdots \mu _{k+1}\parallel \alpha }\right)  \notag \\
&&+B_{\mu _{1}\cdots \mu _{k}}\left( F^{\mu _{1}\cdots \mu _{k}}-\tfrac{%
\left( k+1\right) \left( k+1\right) !}{2\left( a-\left( -\right) ^{k}\frac{%
k+1}{k\left( D-k\right) }\right) }B^{\mu _{1}\cdots \mu _{k}}\right)
\label{ivfo7}
\end{eqnarray}

By construction, the local function (\ref{ivfo7}) is manifestly
gauge-invariant under the gauge transformations (\ref{ivc7}) and (\ref{ivfo6}%
) and, moreover, the elimination of the auxiliary fields $\bar{\omega}%
_{\alpha \parallel \mu _{1}\cdots \mu _{k+1}}$ and $B_{\mu _{1}\cdots \mu
_{k}}$ on their field equations reduces (\ref{ivfo7}) to the second-order
Lagrangian density (\ref{ivc1}). Indeed, the field equations
\begin{eqnarray}
\frac{\delta \overline{\mathcal{L}}_{0}^{\left( IV\right) }}{\delta \bar{%
\omega}^{\alpha \parallel \mu _{1}\cdots \mu _{k+1}}} &\equiv &\tfrac{1}{%
k\left( k+1\right) \left( k+1\right) !}\left( \bar{F}_{\mu _{1}\cdots \mu
_{k+1}\parallel \alpha }-\bar{\omega}_{\left[ \mu _{1}\parallel \mu
_{2}\cdots \mu _{k+1}\right] \alpha }\right) =0,  \label{ivfo8a} \\
\frac{\delta \overline{\mathcal{L}}_{0}^{\left( IV\right) }}{\delta B^{\mu
_{1}\cdots \mu _{k}}} &\equiv &F_{\mu _{1}\cdots \mu _{k}}-\tfrac{\left(
k+1\right) \left( k+1\right) !}{a-\left( -\right) ^{k}\frac{k+1}{k\left(
D-k\right) }}B_{\mu _{1}\cdots \mu _{k}}=0,  \label{ivfo8b}
\end{eqnarray}%
possess the solutions%
\begin{eqnarray}
\bar{\omega}_{\alpha \parallel \mu _{1}\cdots \mu _{k+1}} &=&\left( -\right)
^{k}\left( \bar{F}_{\mu _{1}\cdots \mu _{k+1}\parallel \alpha }-\tfrac{1}{k+1%
}\bar{F}_{\left[ \mu _{1}\cdots \mu _{k+1}\parallel \alpha \right] }\right) ,
\label{ivfo9a} \\
B_{\mu _{1}\cdots \mu _{k}} &=&\tfrac{a-\left( -\right) ^{k}\frac{k+1}{%
k\left( D-k\right) }}{\left( k+1\right) \left( k+1\right) !}F_{\mu
_{1}\cdots \mu _{k}}.  \label{ivfo9b}
\end{eqnarray}%
The Lorentz tensors $\bar{F}_{\mu _{1}\cdots \mu _{k+1}\parallel \alpha }$
that appear in the formulas (\ref{ivfo8a}) and (\ref{ivfo9a}) have the
concrete expressions (\ref{iiifo2}).

Inserting the solutions (\ref{ivfo9a})--(\ref{ivfo9b}) into the first-order
Lagrangian density (\ref{ivfo7}) one finally gets%
\begin{equation}
\mathcal{L}_{0}^{\left( IV\right) }=\left. \overline{\mathcal{L}}%
_{0}^{\left( IV\right) }\right\vert _{\bar{\omega}_{\alpha \parallel \mu
_{1}\cdots \mu _{k+1}}=\left( -\right) ^{k}\left( \bar{F}_{\mu _{1}\cdots
\mu _{k+1}\parallel \alpha }-\tfrac{1}{k+1}\bar{F}_{\left[ \mu _{1}\cdots
\mu _{k+1}\parallel \alpha \right] }\right) ,B_{\mu _{1}\cdots \mu _{k}}=%
\tfrac{a-\left( -\right) ^{k}\frac{k+1}{k\left( D-k\right) }}{\left(
k+1\right) \left( k+1\right) !}F_{\mu _{1}\cdots \mu _{k}}}.  \label{ivfo10}
\end{equation}

\subsection{Case V \label{foV}}

In this situation there are gauge-invariant tensors of $\mathcal{F}_{\mu
_{1}\cdots \mu _{k+1}\alpha }$-type that allows the representation of the
corresponding second-order Lagrangian density quadratically in them. Indeed,
the local function (\ref{vc1}) can be written in terms of the tensors (\ref%
{iiifo2}) as%
\begin{equation}
\mathcal{L}_{0}^{\left( V\right) }=\tfrac{1}{2\left( k+1\right) \left(
k+1\right) !}\bar{F}_{\mu _{1}\cdots \mu _{k+1}\parallel \alpha }\left( -%
\tfrac{\left( -\right) ^{k}}{k+1}\bar{F}^{\mu _{1}\cdots \mu _{k+1}\parallel
\alpha }+\bar{a}\bar{F}^{\mu _{1}\cdots \mu _{k}\alpha \parallel \mu
_{k+1}}\right) .  \label{vfo1}
\end{equation}%
The quantities that play the role of the gauge-invariant tensors are $\bar{F}%
_{\mu _{1}\cdots \mu _{k+1}\parallel \alpha }$ as they verify%
\begin{equation}
\bar{\delta}_{\epsilon ,\varepsilon }^{\left( V\right) }\bar{F}_{\mu
_{1}\cdots \mu _{k+1}\parallel \alpha }=0,  \label{vfo2}
\end{equation}%
under the gauge transformations (\ref{fo2}) corresponding to the analyzed
situation (\ref{vc7}).

By simple algebraic manipulations, the Lagrangian density (\ref{vfo1}) can
be bringed into the form%
\begin{eqnarray}
\mathcal{L}_{0}^{\left( V\right) } &=&\tfrac{\left( -\right) ^{k+1}}{2\left(
k+1\right) ^{3}\left( k+1\right) !}\bar{F}_{\left[ \mu _{1}\cdots \mu
_{k+1}\parallel \alpha \right] }\bar{F}^{\left[ \mu _{1}\cdots \mu
_{k+1}\parallel \alpha \right] }  \notag \\
&&+\tfrac{\left( -\right) ^{k}\left( \bar{a}-\left( -\right) ^{k}\right) }{%
2\left( k+1\right) ^{2}\left( k+1\right) !}\bar{F}_{\mu _{1}\cdots \mu
_{k+1}\parallel \alpha }\bar{F}^{\alpha \left[ \mu _{1}\cdots \mu
_{k}\parallel \mu _{k+1}\right] },  \label{vfo3}
\end{eqnarray}%
that suggests the auxiliary matter fields needed for the linearization
procedure. The first term in the right-hand side of the expression (\ref%
{vfo3}) can be linearized with the help of a matter $\left( k+2\right) $%
-form with coefficients $\bar{B}_{\mu _{1}\cdots \mu _{k+2}}$ [$\bar{B}_{\mu
_{1}\cdots \mu _{k+2}}=\frac{1}{\left( k+2\right) !}\bar{B}_{\left[ \mu
_{1}\cdots \mu _{k+2}\right] }$] while the second through the auxiliary
matter fields $\bar{B}_{\alpha \parallel \mu _{1}\cdots \mu _{k+1}}$ that
satisfy%
\begin{equation}
\bar{B}_{\alpha \parallel \mu _{1}\cdots \mu _{k+1}}=\tfrac{1}{\left(
k+1\right) !}\bar{B}_{\alpha \parallel \left[ \mu _{1}\cdots \mu _{k+1}%
\right] },\quad \sigma ^{\alpha \mu _{1}}\bar{B}_{\alpha \parallel \mu
_{1}\cdots \mu _{k+1}}=0.  \label{vfo4}
\end{equation}

The first-order Lagrangian density associated with (\ref{vc1}) reads as%
\begin{eqnarray}
\overline{\mathcal{L}}_{0}^{\left( V\right) } &\equiv &\overline{\mathcal{L}}%
_{0}^{\left( V\right) }\left( \bar{B}_{\mu _{1}\cdots \mu _{k+2}},\bar{B}%
_{\alpha \parallel \mu _{1}\cdots \mu _{k+1}},F_{\mu _{1}\cdots \mu
_{k+1}\parallel \alpha }\right)  \notag \\
&=&\bar{B}_{\mu _{1}\cdots \mu _{k+2}}\left( \bar{F}^{\left[ \mu _{1}\cdots
\mu _{k+1}\parallel \mu _{k+2}\right] }+\left( -\right) ^{k}\tfrac{\left(
k+1\right) ^{3}\left( k+1\right) !}{2}\bar{B}^{\mu _{1}\cdots \mu
_{k+2}}\right)  \notag \\
&&+\tfrac{\left( -\right) ^{k}-\bar{a}}{\left( 3k+2\right) \left( k+1\right)
\left( k+1\right) !}\bar{B}_{\alpha \parallel \mu _{1}\cdots \mu
_{k+1}}\left( \tfrac{1}{2}\bar{B}^{\alpha \parallel \mu _{1}\cdots \mu
_{k+1}}+\bar{F}^{\mu _{1}\cdots \mu _{k+1}\parallel \alpha }\right.  \notag
\\
&&\left. +\left( -\right) ^{k}\tfrac{k+2}{2\left( 3k+2\right) }\bar{B}^{%
\left[ \mu _{1}\Vert \mu _{2}\cdots \mu _{k+1}\right] \alpha }+\left(
-\right) ^{k}\bar{F}^{\alpha \left[ \mu _{1}\cdots \mu _{k}\parallel \mu
_{k+1}\right] }\right)  \label{vfo5}
\end{eqnarray}%
and this is manifestly gauge-invariant under the gauge transformations (\ref%
{fo2}) [corresponding to the analyzed situation (\ref{vc7})] and%
\begin{equation}
\bar{\delta}_{\epsilon ,\varepsilon }^{\left( V\right) }\bar{B}_{\mu
_{1}\cdots \mu _{k+2}}=0=\bar{\delta}_{\epsilon ,\varepsilon }^{\left(
V\right) }\bar{B}_{\alpha \parallel \mu _{1}\cdots \mu _{k+1}}.  \label{vfo6}
\end{equation}%
Moreover, the elimination of the auxiliary fields in (\ref{vfo5}) on their
own field equations leads to the second-order Lagrangian density (\ref{vc1}%
). Indeed, the field equations%
\begin{eqnarray}
\frac{\delta \overline{\mathcal{L}}_{0}^{\left( V\right) }}{\delta \bar{B}%
^{\mu _{1}\cdots \mu _{k+2}}} &\equiv &\bar{F}_{\left[ \mu _{1}\cdots \mu
_{k+1}\parallel \mu _{k+2}\right] }+\left( -\right) ^{k}\left( k+1\right)
^{3}\left( k+1\right) !\bar{B}_{\mu _{1}\cdots \mu _{k+2}}=0,  \label{vfo7a}
\\
\frac{\delta \overline{\mathcal{L}}_{0}^{\left( V\right) }}{\delta \bar{B}%
^{\alpha \parallel \mu _{1}\cdots \mu _{k+1}}} &\equiv &\tfrac{\left(
-\right) ^{k}-\bar{a}}{\left( 3k+2\right) \left( k+1\right) \left(
k+1\right) !}\left( \bar{B}_{\alpha \parallel \mu _{1}\cdots \mu _{k+1}}+%
\bar{F}_{\mu _{1}\cdots \mu _{k+1}\parallel \alpha }\right.  \notag \\
&&\left. +\left( -\right) ^{k}\tfrac{k+2}{3k+2}\bar{B}_{\left[ \mu _{1}\Vert
\mu _{2}\cdots \mu _{k+1}\right] \alpha }+\left( -\right) ^{k}\bar{F}%
_{\alpha \left[ \mu _{1}\cdots \mu _{k}\parallel \mu _{k+1}\right] }\right)
=0  \label{vfo7b}
\end{eqnarray}%
possess the solutions%
\begin{eqnarray}
\bar{B}_{\mu _{1}\cdots \mu _{k+2}} &=&\tfrac{\left( -\right) ^{k+1}}{\left(
k+1\right) ^{3}\left( k+1\right) !}\bar{F}_{\left[ \mu _{1}\cdots \mu
_{k+1}\parallel \mu _{k+2}\right] },  \label{vfo8a} \\
\bar{B}_{\alpha \parallel \mu _{1}\cdots \mu _{k+1}} &=&-\tfrac{3k+2}{2k+1}%
\left( \bar{F}_{\mu _{1}\cdots \mu _{k+1}\parallel \alpha }+\frac{1}{k}\bar{F%
}_{\left[ \mu _{1}\cdots \mu _{k+1}\parallel \alpha \right] }\right) .
\label{vfo8b}
\end{eqnarray}%
Replacing the solutions (\ref{vfo8a})--(\ref{vfo8b}) into the first-order
Lagrangian density (\ref{vfo5}) one finally reaches to the second-order
Lagrangian density (\ref{vc1})
\begin{equation}
\mathcal{L}_{0}^{\left( V\right) }=\left. \overline{\mathcal{L}}_{0}^{\left(
V\right) }\right\vert _{\bar{B}=\bar{B}\left( \bar{F}\right) }.  \label{vfo9}
\end{equation}

\subsection{Case VI \label{foVI}}

This case mixes, in some sense, the situations analyzed in \ref{foI} and \ref%
{foII}. This is basically due to the expression of the second-oder
Lagrangian density (\ref{vic1}), local function that can be rewritten under
the form%
\begin{equation}
\mathcal{L}_{0}^{\left( VI\right) }\equiv \tfrac{1+\left( -\right) ^{k}k%
\tilde{a}}{k+1}\mathcal{L}_{0}^{\left( I\right) }\left( F_{\mu _{1}\cdots
\mu _{k+1}\parallel \alpha }\right) +\tfrac{\left( 1-\left( -\right) ^{k}%
\tilde{a}\right) k}{k+1}\mathcal{L}_{0}^{\left( II\right) }\left( F_{\mu
_{1}\cdots \mu _{k+1}\parallel \alpha }\right) .  \label{vifo1}
\end{equation}%
The decomposition in the above supplemented with the results obtained in the
subsections \ref{foI} and \ref{foII} allow to linearize the second-oder
Lagrangian density (\ref{vic1}) with the help of two sets of auxiliary
fields. First of them, denoted by $B_{\mu _{1}\cdots \mu _{k+2}}$, is a
matter $\left( k+2\right) $-form [$B_{\mu _{1}\cdots \mu _{k+2}}=\frac{1}{%
\left( k+2\right) !}B_{\left[ \mu _{1}\cdots \mu _{k+2}\right] }$]
\begin{equation}
\bar{\delta}_{\epsilon ,\xi }^{\left( VI\right) }B_{\mu _{1}\cdots \mu
_{k+2}}=0  \label{vifo2}
\end{equation}%
responsible with the linearization of the first term in the right-hand side
of the decomposition (\ref{vifo1}). The second ones, denoted by $\tilde{%
\omega}_{\alpha \parallel \mu _{1}\cdots \mu _{k+1}}$ is a gauge field
\begin{equation}
\bar{\delta}_{\epsilon ,\bar{\epsilon},\xi }^{\left( VI\right) }\tilde{\omega%
}_{\alpha \parallel \mu _{1}\cdots \mu _{k+1}}=\partial _{\alpha }\partial
_{\lbrack \mu _{1}}\bar{\epsilon}_{\mu _{2}\cdots \mu _{k+1}]}  \label{vifo3}
\end{equation}%
antisymmetric in its last two Lorentz indices and is introduced in order to
build the first-order density associated with the second term in the
right-hand side of the decomposition (\ref{vifo1}).

Invoking again the procedures developed in the subsections \ref{foI} and \ref%
{foII} we can write down the first-order Lagrangian density corresponding to
the second-order one (\ref{vifo1})%
\begin{eqnarray}
\overline{\mathcal{L}}_{0}^{\left( VI\right) } &\equiv &\overline{\mathcal{L}%
}_{0}^{\left( VI\right) }\left( B_{\mu _{1}\cdots \mu _{k+2}},\tilde{\omega}%
_{\alpha \parallel \mu _{1}\cdots \mu _{k+1}},F_{\mu _{1}\cdots \mu
_{k+1}\parallel \alpha }\right)  \notag \\
&\equiv &\tfrac{1+\left( -\right) ^{k}k\tilde{a}}{k+1}\mathcal{\bar{L}}%
_{0}^{\left( I\right) }\left( B_{\mu _{1}\cdots \mu _{k+2}},F_{\mu
_{1}\cdots \mu _{k+1}\parallel \alpha }\right) +\tfrac{\left( 1-\left(
-\right) ^{k}\tilde{a}\right) k}{k+1}\mathcal{\bar{L}}_{0}^{\left( II\right)
}\left( \tilde{\omega}_{\alpha \parallel \mu _{1}\cdots \mu _{k+1}},F_{\mu
_{1}\cdots \mu _{k+1}\parallel \alpha }\right) ,  \label{vifo4}
\end{eqnarray}%
where the first term in the right-hand side is given in (\ref{ifo2}) and the
second one has the expression (\ref{iifo6}).

By construction, the Lagrangian density is invariant under the gauge
transformations (\ref{vifo2}), (\ref{vifo3}) and
\begin{equation}
\bar{\delta}_{\epsilon ,\bar{\epsilon},\xi }^{\left( VI\right) }A_{\mu
_{1}\cdots \mu _{k}\parallel \alpha }=\partial _{\lbrack \mu _{1}}\epsilon
_{\mu _{2}\cdots \mu _{k}]\alpha }+\partial _{\lbrack \mu _{1}}\xi _{\mu
_{2}\cdots \mu _{k}]\mid \alpha }+\left( -\right) ^{k}\partial _{\alpha }%
\bar{\epsilon}_{\mu _{1}\cdots \mu _{k}}.  \label{vifo5}
\end{equation}%
The field equations corresponding to the auxiliary variables%
\begin{equation}
\frac{\delta \overline{\mathcal{L}}_{0}^{\left( VI\right) }}{\delta B_{\mu
_{1}\cdots \mu _{k+2}}}=0,\quad \frac{\delta \overline{\mathcal{L}}%
_{0}^{\left( VI\right) }}{\delta \tilde{\omega}^{\alpha \parallel \mu
_{1}\cdots \mu _{k+1}}}=0,  \label{vifo6}
\end{equation}%
have the solutions given in (\ref{ifo5}) and (\ref{iifo9}). Inserting the
aforementioned solutions into the first-order Lagrangian density (\ref{vifo4}%
) one reaches the initial second-order Lagrangian density (\ref{vifo1}).

\subsection{Case VII \label{foVII}}

In this last case, the linearization procedure requires only matter fields.
This is due to the existence of the gauge-invariant tensors of $\mathcal{F}%
_{\mu _{1}\cdots \mu _{k+1}\alpha }$-type that allow the representation of
the second-order Lagrangian (\ref{viic1}) as a bilinear combination in them.
The aforementioned gauge-invariant quantities are exactly the components of
the field-strength $F_{\mu _{1}\cdots \mu _{k+1}\parallel \alpha }$
\begin{equation}
\bar{\delta}_{\epsilon }^{\left( VII\right) }F_{\mu _{1}\cdots \mu
_{k+1}\parallel \alpha }=0,  \label{viifo1}
\end{equation}%
where the gauge transformations are given by (\ref{fo2}) with the right-hand
side expressed by (\ref{3m1}).

In view of constructing the first-order Lagrangian density associated with
the second-order one (\ref{viic1}) we firstly rewrite this as%
\begin{eqnarray}
\mathcal{L}_{0}^{\left( VII\right) } &=&\tfrac{1+\left( -\right) ^{k}ka_{1}}{%
k+1}\mathcal{L}_{0}^{\left( I\right) }\left( F_{\mu _{1}\cdots \mu
_{k+1}\parallel \alpha }\right)  \notag \\
&&+\left( -\right) ^{k}\tfrac{k\left( a_{1}+a_{2}\left( D-k\right) -\left(
-\right) ^{k}\right) }{\left( D-k-1\right) \left( k+1\right) }\mathcal{L}%
_{0}^{\left( II\right) }\left( F_{\mu _{1}\cdots \mu _{k+1}\parallel \alpha
}\right)  \notag \\
&&+\tfrac{k\left( D-k\right) \left( 1-\left( -\right) ^{k}\left(
a_{1}+a_{2}\right) \right) }{\left( D-k-1\right) \left( k+1\right) }\mathcal{%
L}_{0}^{\left( III\right) }\left( F_{\mu _{1}\cdots \mu _{k+1}\parallel
\alpha }\right)  \label{viifo2}
\end{eqnarray}%
and then we make use of the results inferred in the cases \ref{foI}, \ref%
{foII} and \ref{foIII}. The local functions that appear in the right-hand
side of the expression (\ref{viifo2}) have been done in (\ref{ic1}), (\ref%
{iic1}) and (\ref{iiic1}).

The first term in the right-hand side of the decomposition (\ref{viifo2})
can be linearized with the help of the matter $\left( k+2\right) $-form $%
B_{\mu _{1}\cdots \mu _{k+2}}$ as in the situation \ref{foI}, the second one
will be linearized through the antisymmetric matter fields $B_{\alpha \Vert
\mu _{1}\cdots \mu _{k+1}}$ [$B_{\alpha \Vert \mu _{1}\cdots \mu _{k+1}}=%
\frac{1}{\left( k+1\right) !}B_{\alpha \Vert \left[ \mu _{1}\cdots \mu _{k+1}%
\right] }$] and the last one with the help of the traceless matter fields $%
\bar{B}_{\alpha \Vert \mu _{1}\cdots \mu _{k+1}}$ that satisfy the algebraic
properties (\ref{vfo4}). Precisely, the first-order Lagrangian density
associated to the second-order one (\ref{viic1}) reads as%
\begin{eqnarray}
\overline{\mathcal{L}}_{0}^{\left( VII\right) } &\equiv &\overline{\mathcal{L%
}}_{0}^{\left( VII\right) }\left( B_{\mu \nu \rho },B_{\mu \parallel \alpha
\beta },\bar{B}_{\mu \parallel \alpha \beta },F_{\mu \nu \parallel \rho
}\right)  \notag \\
&=&\tfrac{1+\left( -\right) ^{k}ka_{1}}{k+1}\overset{\_}{\mathcal{L}}%
_{0}^{\left( I\right) }\left( B_{\mu _{1}\cdots \mu _{k+2}},F_{\mu
_{1}\cdots \mu _{k+1}\parallel \alpha }\right)  \notag \\
&&+\left( -\right) ^{k}\tfrac{k\left( a_{1}+a_{2}\left( D-k\right) -\left(
-\right) ^{k}\right) }{\left( D-k-1\right) \left( k+1\right) }\overset{\_}{%
\mathcal{L}}_{0}^{\left( II\right) }\left( B_{\alpha \Vert \mu _{1}\cdots
\mu _{k+1}},F_{\mu _{1}\cdots \mu _{k+1}\parallel \alpha }\right)  \notag \\
&&+\tfrac{k\left( D-k\right) \left( 1-\left( -\right) ^{k}\left(
a_{1}+a_{2}\right) \right) }{\left( D-k-1\right) \left( k+1\right) }\overset{%
\_}{\mathcal{L}}_{0}^{\left( III\right) }\left( \bar{B}_{\alpha \Vert \mu
_{1}\cdots \mu _{k+1}},F_{\mu _{1}\cdots \mu _{k+1}\parallel \alpha }\right)
,  \label{viifo3}
\end{eqnarray}%
where the local functions $\overset{\_}{\mathcal{L}}_{0}^{\left( I\right) }$%
, $\overset{\_}{\mathcal{L}}_{0}^{\left( II\right) }$ and $\overset{\_}{%
\mathcal{L}}_{0}^{\left( III\right) }$ are respectively given in (\ref{ifo2}%
), (\ref{iifo6}) and (\ref{iiifo9}). It is obvious that the first-order
Lagrangian density (\ref{viifo3}) is manifestly gauge invariant under the
gauge transformations\
\begin{equation}
\bar{\delta}_{\epsilon }A_{\mu _{1}\cdots \mu _{k}\parallel \alpha
}=\partial _{\lbrack \mu _{1}}\epsilon _{\mu _{2}\cdots \mu _{k}]\parallel
\alpha },\quad \bar{\delta}_{\epsilon }^{\left( VII\right) }B_{\mu
_{1}\cdots \mu _{k+2}}=0,\quad \bar{\delta}_{\epsilon }^{\left( VII\right)
}B_{\alpha \Vert \mu _{1}\cdots \mu _{k+1}}=0=\bar{\delta}_{\epsilon
}^{\left( VII\right) }\bar{B}_{\alpha \Vert \mu _{1}\cdots \mu _{k+1}}
\label{viifo4}
\end{equation}%
and reduces to the second-order Lagrangian (\ref{viic1}) by the elimination
of the auxiliary fields on theirs own field equations.

\section{Conclusions\label{conclRGT}}

In this paper we have given a scheme of unification [at the free level] of a
$\left( k+1\right) $-form and a massless tensor gauge field with the mixed
symmetry $\left( k,1\right) $. The procedure made use of a ($1$-)form-valued
$k$-form [that can be interpreted in terms of a collection of $k$-forms] and
a corresponding Lagrangian density that completely capture the tensor gauge
fields both algebraic and dynamic. Initially, we have constructed the most
general PT-invariant and second-order Lagrangian density that is invariant
under the standard gauge transformations of the just mentioned $k$-forms.
The local function depends on two arbitrary real constants [denoted by $a_{1}
$ and $a_{2}$] and, for some values of the real $a$-parameters, reduces to
standard Lagrangian densities corresponding to the Abelian $\left(
k+1\right) $-form \cite{knaep1} and that of a tensor gauge field with the
mixed symmetry $\left( k,1\right) $ \cite{BB}. Then, we have done the
canonical analysis of the starting Lagrangian theory. This has put into
evidence a partition of the real parameters plane $\left(
a_{1},~a_{2}\right) $ made by seven components. For six among the seven
partition's components it has been shown the generating set of gauge
transformations is \emph{richer} than the original gauge transformations,
including BF \cite{12} and/or conformal-like \cite{CONF}\ gauge
transformations. Moreover, in each of the seven situations has been
calculated the number of independent degrees of freedom and investigated the
presence of unphysical degrees of freedom [ghost-modes]. At this stage, we
have proved the ghost-modes \emph{are absent} only in two of the seven
partition's components namely when the Lagrangian density reduces to that of
a Abelian $\left( k+1\right) $-form and respectively to that of a tensor
gauge field with the mixed symmetry $\left( k,1\right) $, outputs that
generalize the previous results \cite{DESER}. Finally, we have constructed
the first-order formulations associated with the analyzed second-order
Lagrangian theory. Here, for each of the seven partition's components, we
have generated the first-order Lagrangian formulation. These have been done
with the price of adding specific auxiliary gauge/matter fields that made
possible the linearization.

\section*{Acknowledgments}

The author is grateful to Professors Constantin Bizdadea and Solange-Odile
Saliu for useful discussions and comments.


\begin{thebibliography}{99}
\bibitem{ein} A. Einstein, E. G. Straus, \textit{Ann. Math.} \textbf{47},
731 (1946).

\bibitem{sch} E. Schr\"{o}dinger, \textit{Space-Time Structure} (Cambridge
University Press, Cambridge, 1950).

\bibitem{moff} J. W. Moffat,\textit{\ Phys. Rev.} \textit{D} \textbf{19},
3554 (1979).

\bibitem{LP15} E. M. Cioroianu, Int. J. Mod. Phys. \textbf{A27} (2012)
1250189; Rom. J. Phys. \textbf{58} (2013) 529.

\bibitem{knaep1} M. Henneaux, B. Knaepen, C. Schomblond, Commun. Math. Phys.
\textbf{186} (1997) 137

\bibitem{BB} X. Bekaert, N. Boulanger, Commun. Math. Phys. \textbf{245}
(2004) 27; Commun. Math. Phys. \textbf{271} (2007) 723.

\bibitem{DIRAC1} P. A. M. Dirac, \textit{Can. J.Math.} \textbf{2} (1950) 129.

\bibitem{DIRAC2} P. A. M. Dirac, \textit{Lectures on Quantum Mechanics}
(Academic Press, New York, 1967).

\bibitem{1q} M. Henneaux, C. Teitelboim, \textit{Quantization of Gauge
Systems} (Princeton University Press, Princeton, 1992).

\bibitem{12} D. Birmingham, M. Blau, M. Rakowski, G. Thompson, \textit{Phys.
Rept.} \textbf{209} (1991) 129.

\bibitem{CONF} E. S. Fradkin, A. A. Tseytlin, Phys. Rept. \textbf{119}
(1985) 233.

\bibitem{DESER} S. Deser, Gen. Relativ. Gravit. \textbf{1} (1970) 9

\bibitem{zinoviev} Yu. M. Zinoviev, \textit{First Order Formalism for Mixed
Symmetry Tensor Fields}, hep-th/0304067.
\end{thebibliography}
\end{document}